\documentclass[10pt,a4paper]{article}
\usepackage[margin=25mm]{geometry}
\usepackage[utf8]{inputenc}
\usepackage{graphicx}
\usepackage{amsmath,amssymb,amsfonts}
\usepackage{booktabs,array,longtable,tabularx,makecell,threeparttable}
\usepackage{pdflscape,placeins,needspace,textcomp,xurl,xcolor,colortbl}
\usepackage[font=small,labelfont=bf,labelsep=period]{caption}

\DeclareCaptionLabelSeparator{tablebar}{ \textbar{} }
\usepackage[numbers,sort&compress]{natbib}
\usepackage[hidelinks]{hyperref}
\setcitestyle{super,open={},close={}}
\renewcommand{\figurename}{Fig.}
\newcounter{longtableblock}
\begin{document}
\begin{center}
{\LARGE\bfseries Integrating Deep PPG Representations and Digital Biomarkers for Aortic Stenosis Detection with Peri-TAVI Assessment\par}
\end{center}
\noindent Haowan Wang\textsuperscript{1,2}, Cheng Ding\textsuperscript{3}, Donglin Xie\textsuperscript{4}, Huizi Yu\textsuperscript{1}, Guangkun Nie\textsuperscript{4,5,6}, Shenda Hong\textsuperscript{4,7,8,9,*}, Bryan P Yan\textsuperscript{1,*}, Kent Chak-yu So\textsuperscript{1,10,*}.\par\medskip
{\footnotesize
\noindent\textsuperscript{1}Division of Cardiology, Department of Medicine and Therapeutics, The Chinese University of Hong Kong, Hong Kong, China\par
\noindent\textsuperscript{2}Shenzhen Loop Area Institute, Shenzhen, China\par
\noindent\textsuperscript{3}Nanjing University of Aeronautics and Astronautics, Nanjing, China\par
\noindent\textsuperscript{4}National Institute of Health Data Science, Peking University, Beijing, China\par
\noindent\textsuperscript{5}School of Intelligence Science and Technology, Peking University, Beijing, China\par
\noindent\textsuperscript{6}State Key Laboratory of General Artificial Intelligence, Peking University, Beijing, China\par
\noindent\textsuperscript{7}Institute of Medical Technology, Peking University Health Science Center, Beijing, China\par
\noindent\textsuperscript{8}Institute for Artificial Intelligence, Peking University, Beijing, China\par
\noindent\textsuperscript{9}State Key Laboratory of Vascular Homeostasis and Remodeling, NHC Key Laboratory of Cardiovascular Molecular Biology and Regulatory Peptides, Peking University, Beijing, China\par
\noindent\textsuperscript{10}Li Ka Shing Institutes of Health Science, Prince of Wales Hospital, The Chinese University of Hong Kong, Hong Kong, China\par
\noindent\textsuperscript{*}Corresponding authors: Shenda Hong (hongshenda@pku.edu.cn); Bryan P Yan (bryan.yan@cuhk.edu.hk); Kent Chak-yu So (Kentso987@gmail.com).\par}
\section*{Abstract}
\noindent Access to echocardiography is often limited, delaying recognition of progressive aortic stenosis (AS). A brief, low-cost physiological measurement obtained during routine care could help identify patients who may warrant earlier imaging. Here, we present PulseAS, integrating deep photoplethysmography (PPG) representations, waveform biomarkers and clinical information for high-grade AS screening. Three complementary analyses assessed (1) validation and transfer across cohorts differing in geography and ethnic composition, (2) score-based enrichment of AS and pre-transcatheter aortic valve implantation (TAVI) cases, and (3) preoperative and postoperative pulse changes. Using 15-s recordings, separate patient-level cross-validation in EDU (n = 1,209) and MIMIC-III (n = 2,970) yielded AUROCs of 0.818 (PulseAS--XGB) and 0.838 (PulseAS--L2), respectively. In external validation from MIMIC-III to EDU, PulseAS--XGB achieved an AUROC of 0.716, improving to 0.813 after local adaptation. The highest-scoring 10\% captured 45.2\% of all high-grade AS cases in EDU and 41.3\% in MIMIC-III, and 56.9\% of pre-TAVI cases in EDU. Notably, in the EDU TAVI subgroup, PPG waveforms showed earlier systolic peaks and narrower systolic contours following intervention. These findings support clinical--PPG risk stratification in new populations and warrant prospective evaluation of echocardiographic prioritisation.

\section{Introduction}\label{sec:1}

\noindent Aortic stenosis (AS) can remain clinically silent and asymptomatic while pressure overload already drives progressive myocardial hypertrophy, fibrosis and scar, which may persist after valve replacement and worsen long-term outcomes\cite{ESC2025,Everett2018Fibrosis,Thornton2024Scar}. The EARLY TAVR trial supports early transcatheter intervention in asymptomatic severe AS, showing a reduction in the composite of death, stroke or unplanned cardiovascular hospitalization\cite{EARLYTAVR}. RECOVERY and AVATAR, including their long-term follow-up, provide complementary evidence for early surgery in selected patients with asymptomatic severe AS\cite{RECOVERY2020,RECOVERY10Year2026,AVATAR2022,AVATARLongTerm2024}. Early detection of severe AS is therefore increasingly important for timely surveillance and intervention.

However, translating earlier recognition into practice requires more accessible assessment. Echocardiography remains the reference standard for determining valve severity and cardiac consequences, but its dependence on specialized equipment and trained personnel limits routine use, screening, and repeated evaluation\cite{ESC2025}. In Hong Kong, echocardiography waiting times exceeding 68 weeks have been reported as a barrier to diagnosis and monitoring\cite{HAHeartFailure2024}. Similar delays have been reported in Waikato, New Zealand, where urgent echocardiography patients waited 3--4 months and semi-urgent patients about a year\cite{RNZ2025EchoWait}. In low- and middle-income settings, limited equipment and specialist availability restrict access to echocardiography, with advanced assessments often confined to referral centres\cite{GlobalHeart2025Access,CURLS2019}. These constraints highlight the need for more scalable tools that can support risk stratification for high-grade aortic stenosis and prioritise patients for earlier confirmatory echocardiographic assessment.

Accessible physiological signals can extend cardiac assessment beyond imaging. Recent studies have shown that electrocardiograms (ECGs) can detect structural heart disease and hypertrophic cardiomyopathy, highlighting diagnostic information in routinely acquired signals\cite{EchoNext,HCMECGImages}. Photoplethysmography (PPG), recorded by pulse oximeters and wearable devices, captures pulsatile changes in peripheral blood volume\cite{PPGDLReview}. Pulse timing, width and amplitude reflect the interaction of ventricular ejection with the arterial circulation\cite{pyPPG,SVA2024}. The delayed arterial upstroke of AS therefore suggests a quantifiable peripheral pulse expression of valvular obstruction\cite{ASWave2023}. The broader applications in blood-pressure estimation, vascular ageing and cardiovascular screening further demonstrate the physiological information available from optical pulse recordings\cite{SIFPBPNet,SmartwatchHeartAge,PpgAge,MobilePPGOLED}. Whether this information can identify clinically important stenosis and reveal treatment-related changes remains to be determined.

PPG-based studies have begun to establish this potential. PASS evaluated fingertip PPG for AS screening\cite{PASS}, PiLA introduced physiology-guided self-supervised learning for AS\cite{PiLA}, and CardioPPG used ECG-guided representation learning to identify cardiovascular conditions including mitral and aortic valve disease\cite{CardioPPG}. Advances in PPG and multimodal foundation models offer further opportunities to combine learned representations with explicitly defined waveform biomarkers\cite{PaPaGei,PulsePPG,AnyPPG,CSFM}. Yet the clinical value of a pulse-derived score depends on more than classification performance. Because PPG also reflects age, blood pressure, vascular properties and recording conditions\cite{SVA2024,WristSQI}, it is essential to determine what information it adds to routine clinical assessment, whether that information transfers across populations and institutions, and how it relates to observable physiological change. These questions link the development of screening algorithms to the identification of clinically informative peripheral pulse phenotypes.

Valve intervention provides a particularly informative setting in which to examine this physiological relevance. By relieving valvular obstruction and altering ventricular loading, intervention allows pulse morphology to be studied alongside a defined change in cardiovascular physiology\cite{ESC2025,ASWave2023}. A pilot study linking wrist PPG features to invasively measured filling-pressure categories in patients undergoing transcatheter aortic valve implantation (TAVI) has provided initial evidence of a relationship between optical pulse measurements and intracardiac haemodynamics\cite{Scholte}. Comparing recordings before and after intervention can extend this evidence from between-group differences to changes over the treatment course. Together with cross-sectional morphology and external validation, such observations can determine whether PPG captures both disease-related predictive information and a measurable response to intervention—complementary properties of a candidate digital biomarker for screening and longitudinal assessment.

\begin{figure}[!htbp]
\centering

\setbox0=\vbox{\hsize=\linewidth\small\noindent\textbf{\figurename~\thefigure.} PulseAS study overview. \textbf{a,} Proposed clinical pathway and AS analysis cohorts. AS classifiers were developed and internally evaluated separately in EDU and MIMIC-III. \textbf{b,} Disease associated pulse morphology. \textbf{c,} Intervention-based physiological assessment using PPG recordings before and after TAVI. \textbf{d,} External validation and local adaptation of PulseAS--XGB from MIMIC-III to EDU. Medical illustrations and waveforms are schematic.\par}
\includegraphics[width=\linewidth,height=\dimexpr\textheight-\ht0-\dp0-42pt\relax,keepaspectratio]{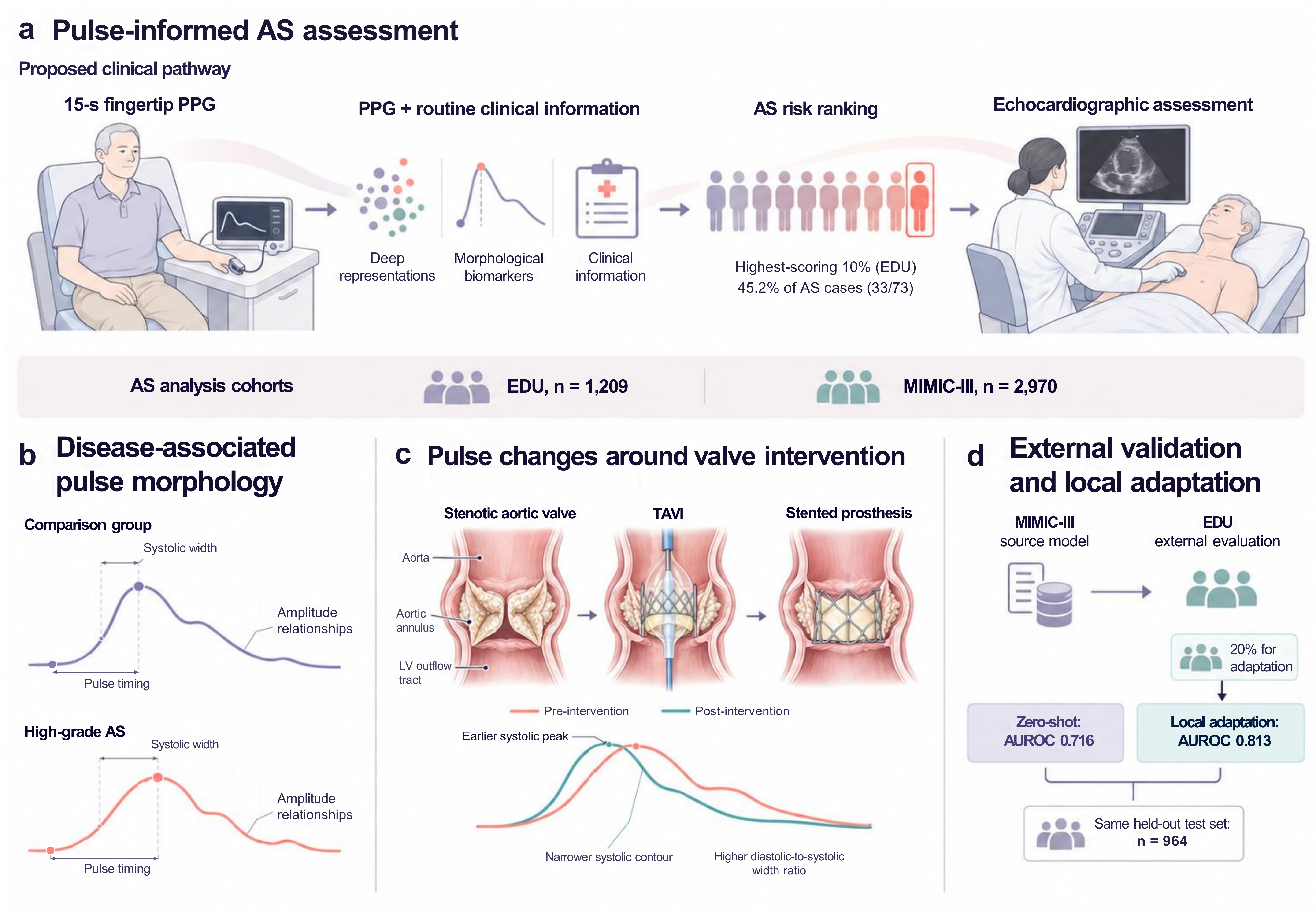}
\caption{PulseAS study overview. \textbf{a,} Proposed clinical pathway and AS analysis cohorts. AS classifiers were developed and internally evaluated separately in EDU and MIMIC-III. \textbf{b,} Disease associated pulse morphology. \textbf{c,} Intervention-based physiological assessment using PPG recordings before and after TAVI. \textbf{d,} External validation and local adaptation of PulseAS--XGB from MIMIC-III to EDU. Medical illustrations and waveforms are schematic.}\label{fig:1}
\end{figure}

Here, we introduce PulseAS, an integrative PPG framework for AS assessment. It combines deep PPG representations, waveform-derived biomarkers and routine clinical information to investigate the peripheral pulse phenotype of high-grade AS (Fig. \ref{fig:1}). The clinical--PPG implementations use XGBoost gradient-boosted trees (PulseAS--XGB), L2-regularised logistic regression (PulseAS--L2) or a multilayer perceptron (PulseAS--MLP). We analyzed 15-s recordings from the Electro-Diagnostic Unit (EDU) cohort and Medical Information Mart for Intensive Care III (MIMIC-III). AS classifiers were developed and internally validated separately within each cohort using patient-level nested cross-validation. In a separate analysis, models developed in MIMIC-III underwent direct external validation in EDU, followed by local adaptation and evaluation on the same held-out test set. High-grade AS, defined as disease more severe than moderate, is the primary screening target.

We show that short-duration PPG captures multidimensional AS-associated pulse patterns and adds predictive information beyond routine clinical variables. Higher-scoring strata concentrate both high-grade AS and pre-TAVI cases, linking the predictions to clinically relevant disease and intervention histories. Crucially, perioperative recordings further reveal earlier systolic peaks, narrower systolic contours and higher diastolic-to-systolic width ratios after valve intervention, providing physiological corroboration alongside disease classification. Discrimination in the external cohort and further improvement after local adaptation extend the predictive evidence beyond the development setting. Together, these findings provide a physiologically supported foundation for scalable PPG-based AS pre-screening and identify peripheral pulse morphology as a candidate measure for repeated assessment after diagnosis. This approach opens a path towards pulse-informed follow-up that could help clinicians recognize patients who warrant earlier echocardiographic reassessment.

\section{Results}\label{sec:2}

\subsection{Study cohorts and participant characteristics}\label{sec:2.1}

\begin{table}[tp]
\centering
\begin{minipage}{\textwidth}
\begingroup
\caption{Clinical characteristics of the study cohorts}\label{tab:1}
\rmfamily\fontsize{8}{10}\selectfont
\setlength{\tabcolsep}{3pt}
\renewcommand{\arraystretch}{1.2}
\arrayrulecolor[gray]{0.85}
\noindent\begin{minipage}{\linewidth}
\noindent\textbf{EDU cohort}\par\smallskip
\begin{tabularx}{\linewidth}{|>{\raggedright\arraybackslash}X|r|r|r|r|}
\hline
\rowcolor[gray]{0.93}
\textbf{Characteristic} & \shortstack[c]{\textbf{Overall}\\n = 1,209} & \shortstack[c]{\textbf{AS $\leq$ moderate}\\n = 1,136} & \shortstack[c]{\textbf{High-grade AS}\\n = 73} & \shortstack[c]{\textbf{P value}}\\
\hline
\rowcolor[gray]{0.965}
\multicolumn{5}{|l|}{\textbf{Demographic and anthropometric characteristics}}\\
\hline
Age, years & 70.0 [60.0, 76.0] & 69.0 [60.0, 76.0] & 76.0 [70.0, 80.0] & \textless{}0.001\\
\hline
Female sex, n/N (\%) & 618/1208 (51.2\%) & 583/1135 (51.4\%) & 35/73 (47.9\%) & 0.629\\
\hline
Height, cm & 160.0 [154.0, 167.0] & 161.0 [154.0, 167.0] & 159.0 [151.0, 165.0] & 0.136\\
\hline
Weight, kg & 63.0 [55.0, 72.0] & 63.0 [55.0, 72.0] & 63.0 [52.1, 70.0] & 0.437\\
\hline
BMI, kg/m\textsuperscript{2} & 24.2 [21.7, 27.3] & 24.2 [21.7, 27.3] & 24.2 [21.6, 27.8] & 0.819\\
\hline
\rowcolor[gray]{0.965}
\multicolumn{5}{|l|}{\textbf{Clinical characteristics}}\\
\hline
SBP, mmHg & 140.0 [127.0, 156.0] & 141.0 [127.0, 156.0] & 134.0 [124.0, 146.2] & 0.013\\
\hline
DBP, mmHg & 76.0 [68.0, 85.0] & 77.0 [69.0, 85.0] & 68.5 [61.0, 75.2] & \textless{}0.001\\
\hline
Documented AF, n/N (\%) & 28/534 (5.2\%) & 15/474 (3.2\%) & 13/60 (21.7\%) & \textless{}0.001\\
\hline
\rowcolor[gray]{0.965}
\multicolumn{5}{|l|}{\textbf{Echocardiographic characteristics}}\\
\hline
LVEF, \% & 57.0 [53.0, 60.0] & 57.0 [53.0, 60.0] & 56.5 [52.5, 59.2] & 0.426\\
\hline
RVSP, mmHg & 27.0 [22.0, 32.3] & 26.8 [22.0, 32.0] & 32.0 [26.0, 38.4] & \textless{}0.001\\
\hline
\end{tabularx}
\end{minipage}\par
\medskip
\noindent\begin{minipage}{\linewidth}
\noindent\textbf{MIMIC-III cohort}\par\smallskip
\begin{tabularx}{\linewidth}{|>{\raggedright\arraybackslash}X|r|r|r|r|}
\hline
\rowcolor[gray]{0.93}
\textbf{Characteristic} & \shortstack[c]{\textbf{Overall}\\n = 2,970} & \shortstack[c]{\textbf{AS $\leq$ moderate}\\n = 2,740} & \shortstack[c]{\textbf{Severe AS}\\n = 230} & \shortstack[c]{\textbf{P value}}\\
\hline
\rowcolor[gray]{0.965}
\multicolumn{5}{|l|}{\textbf{Demographic and anthropometric characteristics}}\\
\hline
Age, years & 66.0 [55.0, 77.0] & 65.0 [54.0, 76.0] & 78.0 [67.2, 84.8] & \textless{}0.001\\
\hline
Female sex, n/N (\%) & 1277/2970 (43.0\%) & 1155/2740 (42.2\%) & 122/230 (53.0\%) & 0.002\\
\hline
Height, cm & 170.2 [162.6, 177.8] & 170.2 [162.6, 177.8] & 167.6 [160.0, 175.3] & 0.010\\
\hline
Weight, kg & 79.4 [66.2, 94.3] & 79.4 [66.7, 94.9] & 77.1 [63.5, 88.8] & 0.038\\
\hline
BMI, kg/m\textsuperscript{2} & 27.4 [23.7, 31.9] & 27.4 [23.7, 31.9] & 27.2 [23.6, 31.3] & 0.480\\
\hline
\rowcolor[gray]{0.965}
\multicolumn{5}{|l|}{\textbf{Clinical characteristics}}\\
\hline
SBP, mmHg & 121.0 [106.0, 137.0] & 121.0 [106.0, 137.0] & 121.5 [109.0, 135.0] & 0.716\\
\hline
DBP, mmHg & 65.0 [56.0, 75.0] & 65.0 [56.0, 76.0] & 62.0 [53.0, 70.0] & \textless{}0.001\\
\hline
Heart rate, beats/min & 78.0 [67.0, 92.0] & 79.0 [68.0, 92.0] & 75.0 [65.0, 85.0] & 0.002\\
\hline
AF during index echocardiography admission, n/N (\%) & 763/2436 (31.3\%) & 666/2251 (29.6\%) & 97/185 (52.4\%) & \textless{}0.001\\
\hline
\rowcolor[gray]{0.965}
\multicolumn{5}{|l|}{\textbf{Echocardiographic characteristics}}\\
\hline
LVEF, \% & 47.5 [35.0, 55.0] & 47.5 [35.0, 60.0] & 45.0 [35.0, 52.5] & 0.254\\
\hline
\rowcolor[gray]{0.965}
\multicolumn{5}{|l|}{\textbf{Recorded ethnicity, n/N (\%)}}\\
\hline
White & 2172/2970 (73.1\%) & 1973/2740 (72.0\%) & 199/230 (86.5\%) & \\
\hline
Black & 304/2970 (10.2\%) & 299/2740 (10.9\%) & 5/230 (2.2\%) & \\
\hline
Hispanic & 104/2970 (3.5\%) & 98/2740 (3.6\%) & 6/230 (2.6\%) & \\
\hline
Asian & 82/2970 (2.8\%) & 81/2740 (3.0\%) & 1/230 (0.4\%) & \\
\hline
Native & 2/2970 (0.1\%) & 2/2740 (0.1\%) & 0/230 (0.0\%) & \\
\hline
Other & 96/2970 (3.2\%) & 89/2740 (3.2\%) & 7/230 (3.0\%) & \\
\hline
Unknown & 210/2970 (7.1\%) & 198/2740 (7.2\%) & 12/230 (5.2\%) & \\
\hline
\end{tabularx}
\par\smallskip

\end{minipage}\par
\endgroup

\end{minipage}

\end{table}

\begin{figure}[!htb]
\centering

\setbox0=\vbox{\hsize=\linewidth\small\noindent\textbf{\figurename~\thefigure.} AS cohort selection. \textbf{a,b,} EDU and MIMIC-III, respectively. Counts represent patients, with mutually exclusive exclusions.\par}
\includegraphics[width=\linewidth,height=\dimexpr\textheight-\ht0-\dp0-42pt\relax,keepaspectratio]{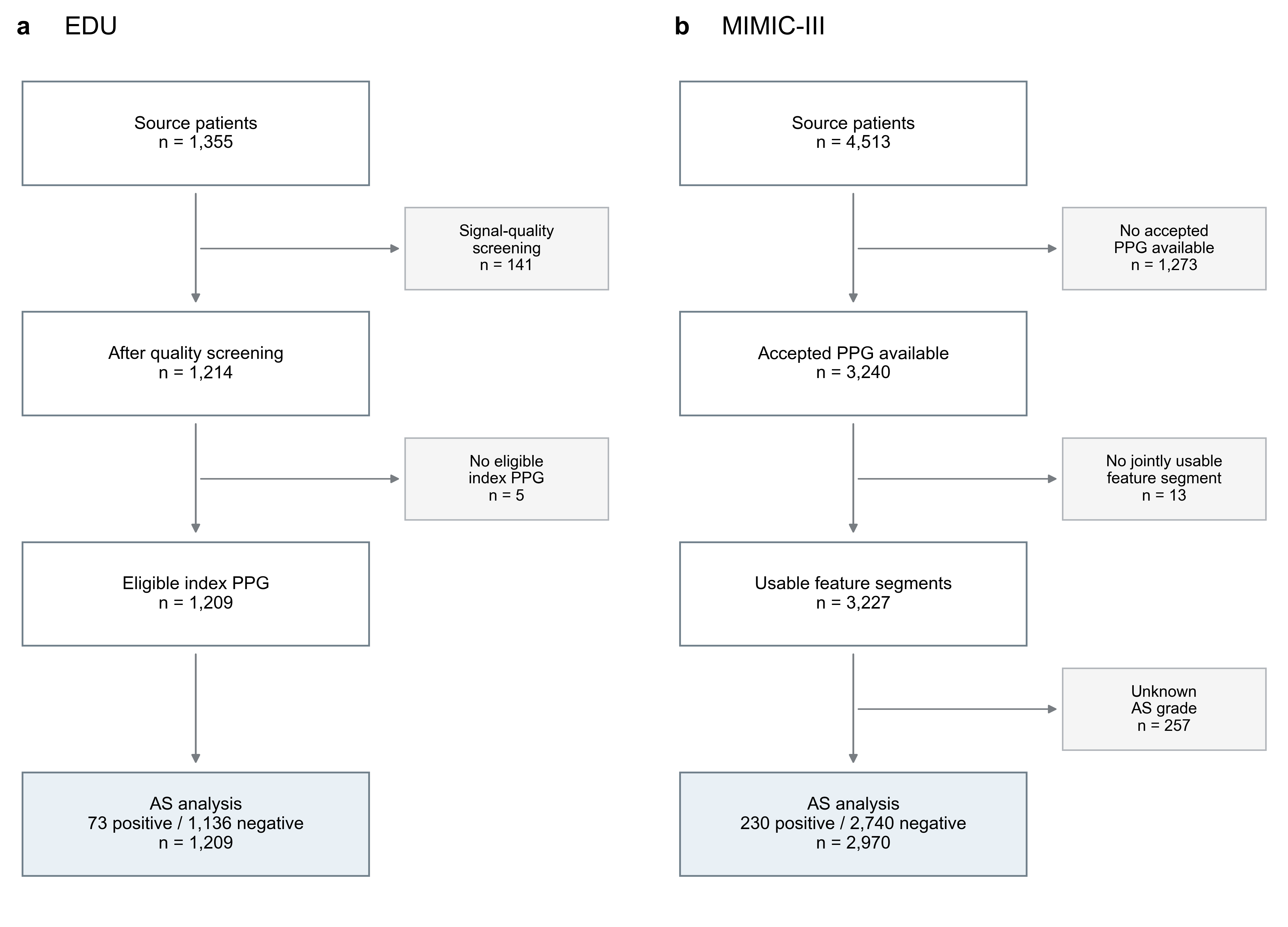}
\caption{AS cohort selection. \textbf{a,b,} EDU and MIMIC-III, respectively. Counts represent patients, with mutually exclusive exclusions.}\label{fig:2}
\end{figure}

The study comprised two clinically distinct populations: patients attending echocardiography in Hong Kong (EDU) and patients receiving critical care in Boston (MIMIC-III; Fig. \ref{fig:1}). The analysis included 1,209 EDU patients, of whom 73 had high-grade AS (10 moderate-to-severe and 63 severe), and 2,970 MIMIC-III patients, including 230 with severe AS. Comparison patients had AS no greater than moderate, with concomitant valve disease retained in the primary analyses (Table \ref{tab:1}; Fig. \ref{fig:2}).

The cohorts differed in demographic and clinical composition. EDU was predominantly Asian, whereas 73.1\% of MIMIC-III patients were recorded as White. Despite these differences, AS-positive patients in both cohorts were older and had lower diastolic blood pressure (DBP) than comparison patients. Lower systolic blood pressure (SBP) was also observed in EDU. Documented atrial fibrillation (AF) was more frequent in both AS groups among patients with available records (Table \ref{tab:1}). Other clinical characteristics included body mass index (BMI), left ventricular ejection fraction (LVEF) and right ventricular systolic pressure (RVSP).

Longitudinal recordings from 57 EDU patients with AS enabled assessment of whether disease-associated pulse features changed after valve intervention. These comprised 53 preoperative, 40 postoperative and 13 follow-up assessments, including 38 patients with paired preoperative and postoperative recordings.

\subsection{PPG reveals delayed systolic pulse development in AS}\label{sec:2.2}

\begin{figure}[!htbp]
\centering

\setbox0=\vbox{\hsize=\linewidth\small\noindent\textbf{\figurename~\thefigure.} AS-associated pulse phenotype and physiological changes after TAVI. \textbf{a,b,} Median PPG waveforms by AS group. \textbf{c,} Paired preoperative and postoperative waveforms. \textbf{d--f,} Corresponding second derivatives. Shading denotes pointwise 95\% percentile bootstrap CIs. \textbf{g,} Directions of AS-associated morphology differences and changes after TAVI. \textbf{h,} Feature distributions across reference, preoperative, postoperative and follow-up assessments. The reference group had no more than mild disease across assessed valve lesions. Points denote patient-stage observations, lines connect repeated measurements, and bars indicate medians and interquartile ranges. Feature comparisons include all available preoperative and postoperative observations. q values are Benjamini--Hochberg-adjusted across 510 features per contrast (*q \textless{} 0.05, **q \textless{} 0.01, ***q \textless{} 0.001; ns, q \ensuremath{\geq} 0.05).\par}
\includegraphics[width=\linewidth,height=\dimexpr\textheight-\ht0-\dp0-42pt\relax,keepaspectratio]{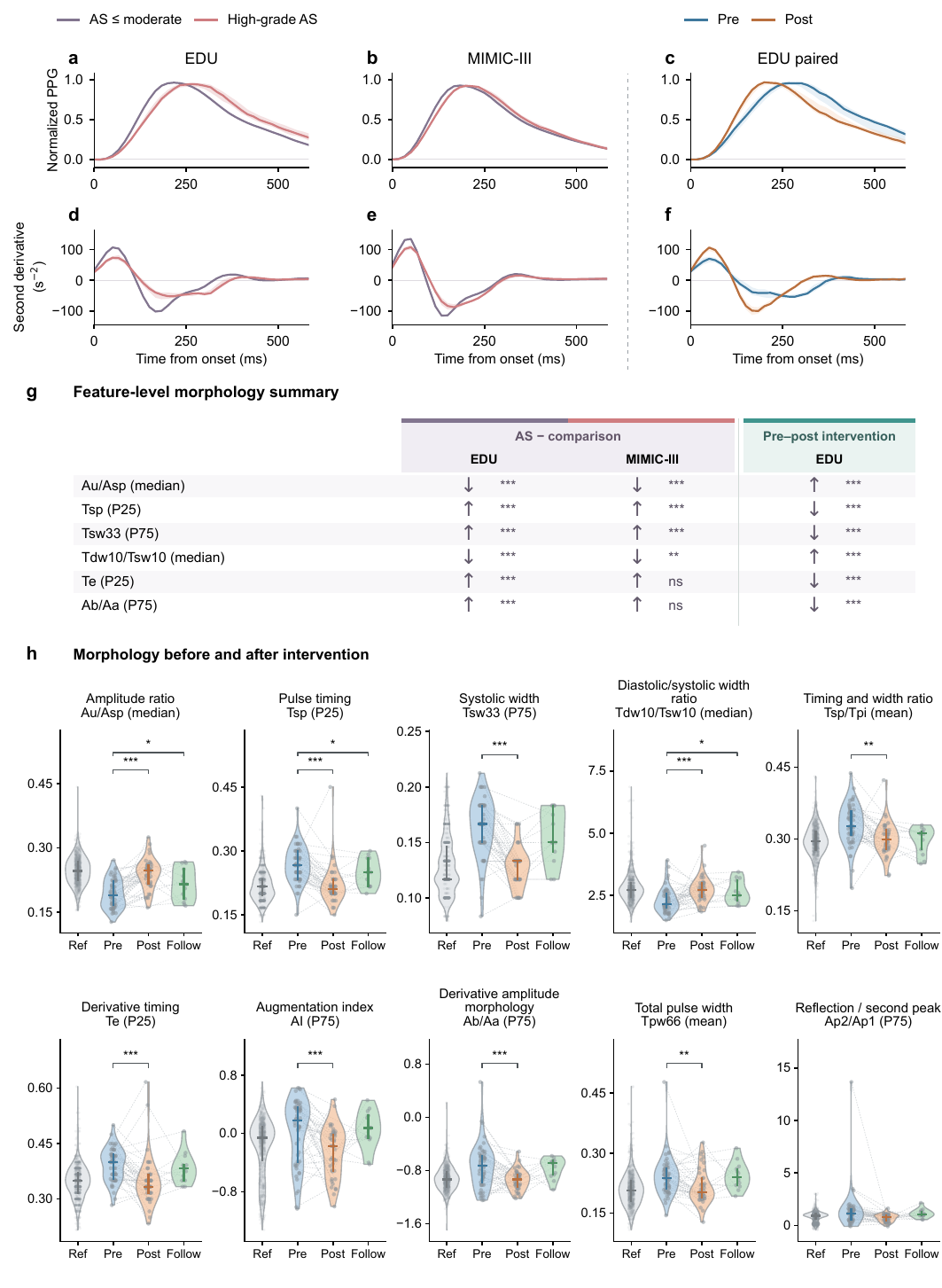}
\caption{AS-associated pulse phenotype and physiological changes after TAVI. \textbf{a,b,} Median PPG waveforms by AS group. \textbf{c,} Paired preoperative and postoperative waveforms. \textbf{d--f,} Corresponding second derivatives. Shading denotes pointwise 95\% percentile bootstrap CIs. \textbf{g,} Directions of AS-associated morphology differences and changes after TAVI. \textbf{h,} Feature distributions across reference, preoperative, postoperative and follow-up assessments. The reference group had no more than mild disease across assessed valve lesions. Points denote patient-stage observations, lines connect repeated measurements, and bars indicate medians and interquartile ranges. Feature comparisons include all available preoperative and postoperative observations. q values are Benjamini--Hochberg-adjusted across 510 features per contrast (*q \textless{} 0.05, **q \textless{} 0.01, ***q \textless{} 0.001; ns, q \ensuremath{\geq} 0.05).}
\label{fig:3}
\end{figure}

\noindent Population waveforms and quantitative morphology features provided complementary descriptions of the AS-associated pulse phenotype. As shown in Fig. \ref{fig:3}a,b, AS-positive groups had later systolic peaks, with more pronounced waveform separation in EDU than in MIMIC-III. The waveform landmarks underlying the representative morphology features are illustrated in Supplementary Fig. \ref{fig:S1}. Waveform summaries are shown with pointwise 95\% confidence intervals (CIs). Feature-level comparisons identified concordant increases in onset to-peak time, Tsp, and systolic width, Tsw33, together with decreases in Au/Asp and Tdw10/Tsw10 (Fig. \ref{fig:3}g). These changes jointly described delayed systolic peaking and altered amplitude and pulse-width relationships. Across the full morphology set, 284 of 510 features differed in EDU and 282 in MIMIC-III after multiplicity correction (q \ensuremath{<} 0.05; Supplementary Fig. \ref{fig:S3}).

To test whether the AS-associated pulse phenotype changed after valve intervention, we compared preoperative and postoperative recordings. In paired recordings, postoperative median pulses peaked earlier and had narrower systolic contours (Fig. \ref{fig:3}c,f). The feature-level analysis likewise showed decreases in Tsp and Tsw33 and increases in Au/Asp and Tdw10/Tsw10, consistent with attenuation of AS-associated pulse abnormalities. Across available observations, preoperative and postoperative medians were 0.267 and 0.210 s for Tsp (25th percentile, P25), 0.167 and 0.133 s for Tsw33 (75th percentile, P75), and 2.151 and 2.707 for Tdw10/Tsw10 (median), respectively. Tsp/Tpi also decreased, indicating earlier peaking relative to the pulse interval (Fig. \ref{fig:3}h).

Nine of the ten representative longitudinal features differed between preoperative and postoperative stages in the full-data analysis (all q \textless{} 0.01). At follow-up, Au/Asp, Tsp and Tdw10/Tsw10 remained different from preoperative values (all q \textless{} 0.05), while no postoperative-to-follow-up differences were detected among the ten features. The attenuation of AS-associated pulse features after TAVI provides intervention-based physiological support for the disease-associated phenotype.

\subsection{Clinical and PPG data discriminate AS across patient groups}\label{sec:2.3}

\begin{figure}[!htbp]
\centering

\setbox0=\vbox{\hsize=\linewidth\small\noindent\textbf{\figurename~\thefigure.} AS discrimination, incremental value of PPG and performance across clinical strata. EDU results are shown in the left two columns and MIMIC-III results in the right two columns. \textbf{a,c,} Representative clinical (L2 logistic regression), morphology (full pyPPG--ExtraTrees), frozen AnyPPG (XGBoost) and PulseAS strategies in EDU and MIMIC-III. PulseAS--XGB and PulseAS--L2 were selected in the respective cohorts. \textbf{b,d,} Matched L2 comparisons of Clinical-4 (age, sex, SBP and DBP) with and without PPG. \textbf{e,f,i,j} and \textbf{g,h,k,l,} Age, sex, SBP (\textless{}130/\ensuremath{\geq}130 mmHg) and DBP (\textless{}80/\ensuremath{\geq}80 mmHg) strata in EDU and MIMIC-III, respectively, using the selected PulseAS models. Dashed diagonals indicate chance discrimination.\par}
\includegraphics[width=\linewidth,height=\dimexpr\textheight-\ht0-\dp0-42pt\relax,keepaspectratio]{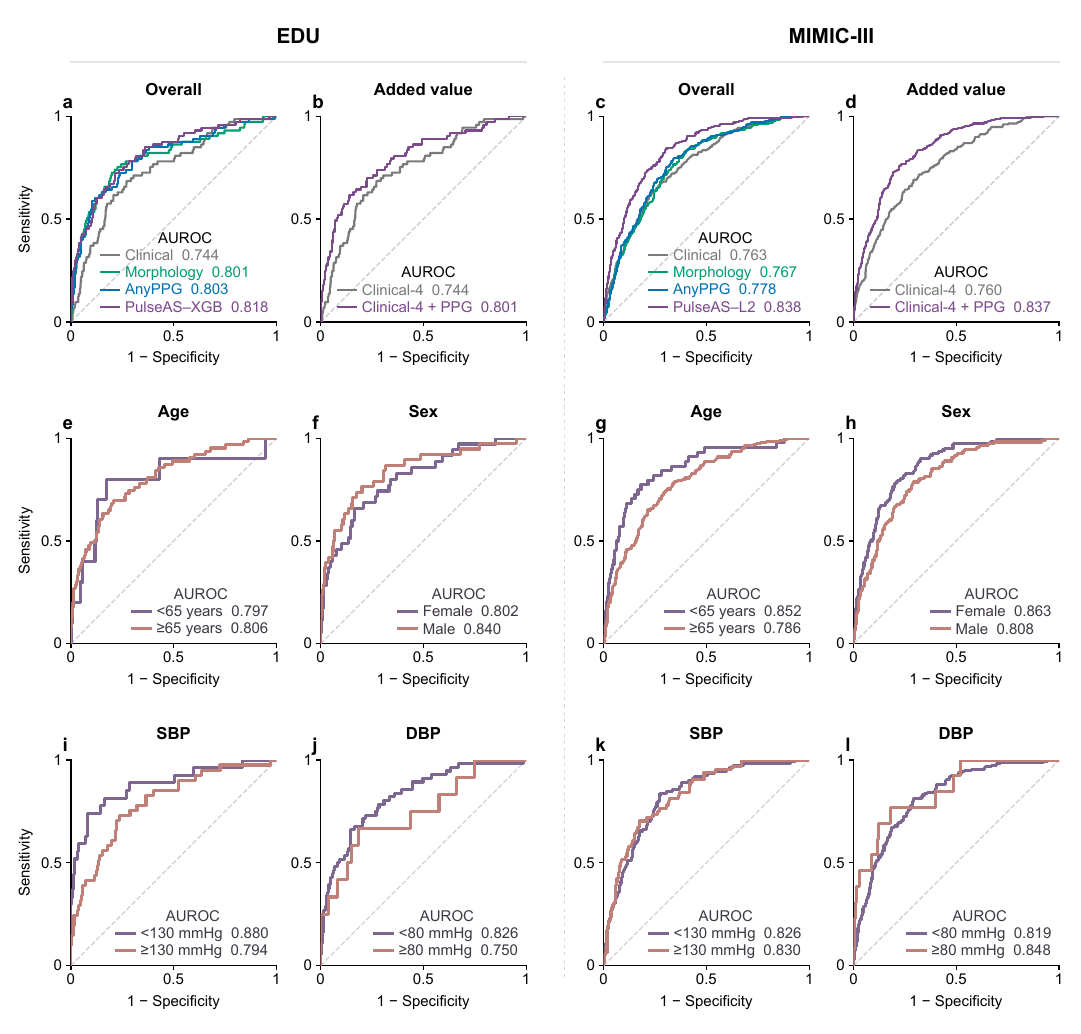}
\caption{AS discrimination, incremental value of PPG and performance across clinical strata. EDU results are shown in the left two columns and MIMIC-III results in the right two columns. \textbf{a,c,} Representative clinical (L2 logistic regression), morphology (full pyPPG--ExtraTrees), frozen AnyPPG (XGBoost) and PulseAS strategies in EDU and MIMIC-III. PulseAS--XGB and PulseAS--L2 were selected in the respective cohorts. \textbf{b,d,} Matched L2 comparisons of Clinical-4 (age, sex, SBP and DBP) with and without PPG. \textbf{e,f,i,j} and \textbf{g,h,k,l,} Age, sex, SBP (\textless{}130/\ensuremath{\geq}130 mmHg) and DBP (\textless{}80/\ensuremath{\geq}80 mmHg) strata in EDU and MIMIC-III, respectively, using the selected PulseAS models. Dashed diagonals indicate chance discrimination.}\label{fig:4}
\end{figure}

\noindent To assess the predictive relevance of these pulse patterns, we compared morphology-based models, pretrained PPG representations and their combinations with clinical variables with classifiers developed and internally evaluated separately in each cohort using patient-level nested cross-validation (Table \ref{tab:2}). Both forms of PPG information supported discrimination: morphology-based models achieved areas under the receiver operating characteristic curve (AUROCs) of 0.801 in EDU and 0.767 in MIMIC-III, while models using frozen AnyPPG representations achieved 0.803 and 0.778, respectively (Fig. \ref{fig:4}a,c).

Combining clinical and PPG information yielded the highest AUROC and average precision (AP) point estimates among the eight strategies. In EDU, PulseAS--XGB achieved an AUROC of 0.818 (95\% CI, 0.763--0.868) and an AP of 0.356. In MIMIC-III, PulseAS--L2 achieved an AUROC of 0.838 (0.813--0.863) and an AP of 0.312. Sensitivities at the training-derived screening thresholds were 0.863 and 0.896, respectively. These models formed the basis of the subgroup and score-stratification analyses.

Across age, sex and recorded blood-pressure strata, the selected models retained discrimination, with AUROC point estimates ranging from 0.750 to 0.880 in EDU (Fig. \ref{fig:4}e,f,i,j) and from 0.786 to 0.863 in MIMIC-III (Fig. \ref{fig:4}g,h,k,l). Finer age stratification revealed no monotonic pattern in EDU and decreasing AUROC point estimates in MIMIC-III. The youngest MIMIC-III group contained only two AS cases (Supplementary Fig. \ref{fig:S4}).

AS discrimination was higher when comparison patients had no more than mild disease in every assessed valve lesion. In this sensitivity analysis, AUROC increased from 0.818 to 0.834 for PulseAS--XGB in EDU and from 0.838 to 0.897 for PulseAS--L2 in MIMIC-III (Supplementary Table \ref{tab:S1}).

\begin{table}[!htbp]
\caption{Internal validation of AS classifiers}\label{tab:2}
\centering
\begingroup
\rmfamily\fontsize{8}{10.5}\selectfont
\setlength{\tabcolsep}{1.5pt}
\renewcommand{\arraystretch}{1.15}
\begin{tabular*}{\linewidth}{@{\extracolsep{\fill}}>{\raggedright\arraybackslash}m{94pt}rrrrr@{}}
\toprule
Model & \shortstack[r]{AUROC\\(95\% CI)} & \shortstack[r]{AP\\(95\% CI)} & \shortstack[r]{Brier\\(95\% CI)} & \shortstack[r]{Sens@\\60\%Spec} & \shortstack[r]{Balanced\\accuracy}\\
\midrule
\multicolumn{6}{@{}l}{\textbf{EDU}}\\*
Full pyPPG + ExtraTrees & \shortstack[r]{0.801\\(0.738--0.860)} & \shortstack[r]{0.300\\(0.209--0.421)} & \shortstack[r]{0.050\\(0.041--0.060)} & 0.822 & 0.723\\[4pt]
Clinical + Logistic L2 & \shortstack[r]{0.744\\(0.682--0.802)} & \shortstack[r]{0.155\\(0.112--0.223)} & \shortstack[r]{0.054\\(0.043--0.065)} & 0.712 & 0.680\\[4pt]
Frozen AnyPPG + XGBoost & \shortstack[r]{0.803\\(0.745--0.854)} & \shortstack[r]{0.285\\(0.196--0.409)} & \shortstack[r]{0.050\\(0.041--0.060)} & 0.849 & 0.688\\[4pt]
AnyPPG--Top-10 pyPPG fusion + XGBoost & \shortstack[r]{0.793\\(0.735--0.846)} & \shortstack[r]{0.259\\(0.184--0.378)} & \shortstack[r]{0.051\\(0.042--0.061)} & 0.863 & 0.682\\[4pt]
AnyPPG--Top-10 neural fusion & \shortstack[r]{0.740\\(0.669--0.806)} & \shortstack[r]{0.271\\(0.181--0.375)} & \shortstack[r]{0.051\\(0.042--0.062)} & 0.822 & 0.693\\[4pt]
\textbf{PulseAS--XGB} & \shortstack[r]{\textbf{0.818}\\\textbf{(0.763--0.868)}} & \shortstack[r]{\textbf{0.356}\\\textbf{(0.253--0.475)}} & \shortstack[r]{\textbf{0.048}\\\textbf{(0.039--0.058)}} & \textbf{0.863} & \textbf{0.700}\\[4pt]
PulseAS--MLP & \shortstack[r]{0.758\\(0.689--0.823)} & \shortstack[r]{0.289\\(0.197--0.397)} & \shortstack[r]{0.051\\(0.041--0.061)} & 0.781 & 0.695\\[4pt]
PulseAS--L2 & \shortstack[r]{0.801\\(0.742--0.852)} & \shortstack[r]{0.324\\(0.223--0.434)} & \shortstack[r]{0.049\\(0.040--0.059)} & 0.822 & 0.752\\[4pt]
\arrayrulecolor{black}\midrule[0.5pt]\arrayrulecolor[gray]{0.85}
\multicolumn{6}{@{}l}{\textbf{MIMIC-III}}\\*
Full pyPPG + ExtraTrees & \shortstack[r]{0.767\\(0.737--0.798)} & \shortstack[r]{0.203\\(0.174--0.247)} & \shortstack[r]{0.067\\(0.065--0.069)} & 0.826 & 0.712\\[4pt]
Clinical + Logistic L2 & \shortstack[r]{0.763\\(0.733--0.793)} & \shortstack[r]{0.243\\(0.201--0.299)} & \shortstack[r]{0.066\\(0.064--0.067)} & 0.774 & 0.688\\[4pt]
Frozen AnyPPG + XGBoost & \shortstack[r]{0.778\\(0.750--0.806)} & \shortstack[r]{0.207\\(0.180--0.251)} & \shortstack[r]{0.066\\(0.064--0.068)} & 0.857 & 0.712\\[4pt]
AnyPPG--Top-10 pyPPG fusion + XGBoost & \shortstack[r]{0.780\\(0.753--0.807)} & \shortstack[r]{0.210\\(0.182--0.255)} & \shortstack[r]{0.066\\(0.064--0.068)} & 0.839 & 0.709\\[4pt]
AnyPPG--Top-10 neural fusion & \shortstack[r]{0.778\\(0.751--0.808)} & \shortstack[r]{0.212\\(0.183--0.254)} & \shortstack[r]{0.066\\(0.064--0.067)} & 0.861 & 0.709\\[4pt]
PulseAS--XGB & \shortstack[r]{0.832\\(0.807--0.855)} & \shortstack[r]{0.296\\(0.253--0.352)} & \shortstack[r]{0.062\\(0.060--0.064)} & 0.913 & 0.748\\[4pt]
PulseAS--MLP & \shortstack[r]{0.820\\(0.795--0.847)} & \shortstack[r]{0.297\\(0.256--0.360)} & \shortstack[r]{0.062\\(0.060--0.064)} & 0.883 & 0.725\\[4pt]
\textbf{PulseAS--L2} & \shortstack[r]{\textbf{0.838}\\\textbf{(0.813--0.863)}} & \shortstack[r]{\textbf{0.312}\\\textbf{(0.272--0.375)}} & \shortstack[r]{\textbf{0.061}\\\textbf{(0.059--0.063)}} & \textbf{0.896} & \textbf{0.748}\\[4pt]
\bottomrule
\end{tabular*}
\endgroup
\par\smallskip
{\fontsize{8}{10}\selectfont\noindent Bold rows indicate the PulseAS implementation with the highest internal-validation AUROC in each cohort, selected for subsequent subgroup and score-stratification analyses.\par}
\end{table}

\subsection{Incremental value of PPG beyond clinical predictors}\label{sec:2.4}

\noindent Matched L2 comparisons quantified the contribution of PPG beyond age, sex and systolic and diastolic blood pressure while holding the classifier, patients and validation folds constant. Adding PPG increased AUROC from 0.744 to 0.801 in EDU and from 0.760 to 0.837 in MIMIC-III, corresponding to increments of 0.057 and 0.077, respectively (Fig. \ref{fig:4}b,d; Supplementary Table \ref{tab:S2}).

This direction of change extended across all five clinical baselines incorporating blood pressure with or without age and sex. For pulse pressure alone, adding PPG increased AUROC from 0.578 to 0.767 in EDU and from 0.640 to 0.811 in MIMIC-III. The direction of change was preserved after complete-case refitting and allowance for nonlinear clinical effects, with AUROC point estimates favouring the inclusion of PPG in both settings. In MIMIC-III, the increments beyond the four-variable baseline were 0.092 after complete-case refitting and 0.089 with nonlinear clinical terms. The corresponding EDU estimates were 0.068 and 0.076. Paired estimates and confidence intervals are reported in Supplementary Table \ref{tab:S2}. These comparisons support PPG as a complementary source of predictive information alongside routinely measured clinical variables.

\subsection{Score-based stratification enriches for AS and TAVI cases}\label{sec:2.5}

\noindent The selected PulseAS models also yielded lower overall probability error than the clinical baselines, with Brier scores of 0.048 versus 0.054 in EDU and 0.061 versus 0.066 in MIMIC-III (Table \ref{tab:2}). Calibration curves are presented in Supplementary Fig. \ref{fig:S2}.

\begin{figure}[!htbp]
\centering

\setbox0=\vbox{\hsize=\linewidth\small\noindent\textbf{\figurename~\thefigure.} AS case capture and enrichment by PulseAS score. \textbf{a,b,} EDU. \textbf{c,d,} MIMIC-III. Left panels show cumulative case capture; right panels show enrichment. Case capture denotes the proportion of all cases in the corresponding cohort included in the selected score stratum. Enrichment is the included-group case proportion divided by cohort prevalence. Shading indicates pointwise 95\% CIs; dashed reference lines indicate random capture or no enrichment.\par}
\includegraphics[width=\linewidth,height=\dimexpr\textheight-\ht0-\dp0-42pt\relax,keepaspectratio]{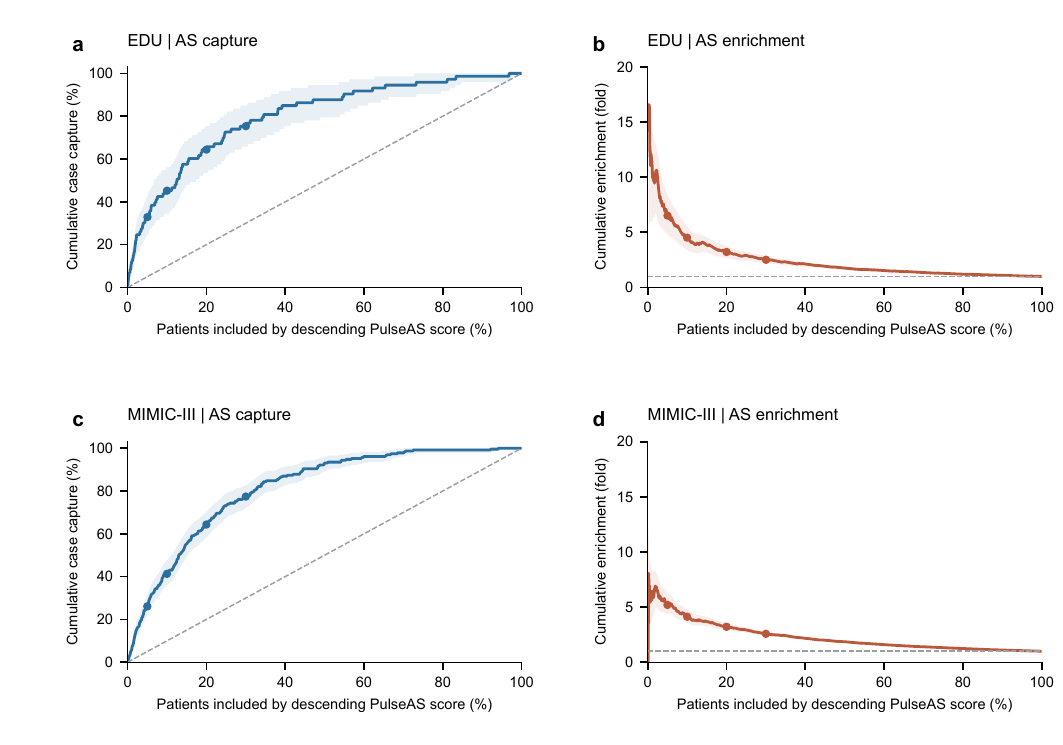}
\caption{AS case capture and enrichment by PulseAS score. \textbf{a,b,} EDU. \textbf{c,d,} MIMIC-III. Left panels show cumulative case capture; right panels show enrichment. Case capture denotes the proportion of all cases in the corresponding cohort included in the selected score stratum. Enrichment is the included-group case proportion divided by cohort prevalence. Shading indicates pointwise 95\% CIs; dashed reference lines indicate random capture or no enrichment.}\label{fig:5}
\end{figure}

Ranking patients within each cohort by their calibrated out-of-fold (OOF) PulseAS scores concentrated high-grade AS cases within the highest-scoring strata (Fig. \ref{fig:5}). In EDU, the highest scoring 10\% comprised 121 patients and included 33 of 73 high-grade AS cases, capturing 45.2\% of cases. In MIMIC-III, the corresponding 297 patients included 95 of 230 cases, giving 41.3\% case capture. AS proportions in these strata were 27.3\% and 32.0\%, compared with overall prevalences of 6.0\% and 7.7\%, corresponding to 4.52-fold and 4.13-fold enrichment. Expanding the selected fraction to 20\% captured 64.4\% and 64.3\% of cases, respectively (Table \ref{tab:3}).

Linkage to procedure records showed that the EDU rankings also concentrated pre-TAVI cases, although pre-TAVI was neither a model input nor a prediction target. The highest-scoring 10\% included 29 of 51 pre-TAVI cases (56.9\%). Pre-TAVI cases accounted for 24.0\% of this stratum, compared with 4.2\% overall, corresponding to 5.68-fold enrichment. The highest-scoring 20\% included 40 cases (78.4\%; Table \ref{tab:3}). The score-defined strata therefore concentrated both high-grade AS and pre-TAVI patients, connecting disease classification with an additional dimension of clinical relevance.

\begin{table}[!htbp]
\centering
\setcounter{table}{2}
\renewcommand{\theHtable}{main.\arabic{table}}
\caption{AS and pre-TAVI across PulseAS score strata}
\label{tab:3}
\fontsize{9}{11}\selectfont
\setlength{\tabcolsep}{2pt}
\renewcommand{\arraystretch}{1.3}
\begin{tabular}{@{}>{\raggedright\arraybackslash}p{50.92356pt}
>{\raggedleft\arraybackslash}p{47.44161pt}
>{\raggedleft\arraybackslash}p{91.83651pt}
>{\raggedleft\arraybackslash}p{73.12101pt}
>{\raggedleft\arraybackslash}p{77.90870pt}
>{\raggedleft\arraybackslash}p{94.01273pt}@{}}
\toprule
\multicolumn{1}{@{}l}{\makecell[l]{Highest-\\scoring\\fraction}} &
\multicolumn{1}{c}{\makecell[c]{Patients}} &
\multicolumn{1}{c}{\makecell[c]{AS capture\\n/N (\%)}} &
\multicolumn{1}{c}{\makecell[c]{AS proportion\\\%}} &
\multicolumn{1}{c}{\makecell[c]{AS enrichment\\fold}} &
\multicolumn{1}{c@{}}{\makecell[c]{Pre-TAVI capture\\n/N (\%)}} \\
\midrule
\multicolumn{6}{@{}l}{\textbf{EDU}} \\[2pt]
5\% & 61 & 24/73 (32.9\%) & 39.3 & 6.52 & 21/51 (41.2\%)\\
10\% & 121 & 33/73 (45.2\%) & 27.3 & 4.52 & 29/51 (56.9\%)\\
20\% & 242 & 47/73 (64.4\%) & 19.4 & 3.22 & 40/51 (78.4\%)\\
30\% & 363 & 55/73 (75.3\%) & 15.2 & 2.51 & 46/51 (90.2\%)\\
50\% & 605 & 64/73 (87.7\%) & 10.6 & 1.75 & 48/51 (94.1\%)\\
100\% & 1,209 & 73/73 (100.0\%) & 6.0 & 1.00 & 51/51 (100.0\%)\\
\midrule
\multicolumn{6}{@{}l}{\textbf{MIMIC-III}} \\[2pt]
5\% & 149 & 60/230 (26.1\%) & 40.3 & 5.20 & /\\
10\% & 297 & 95/230 (41.3\%) & 32.0 & 4.13 & /\\
20\% & 594 & 148/230 (64.3\%) & 24.9 & 3.22 & /\\
30\% & 891 & 178/230 (77.4\%) & 20.0 & 2.58 & /\\
50\% & 1,485 & 214/230 (93.0\%) & 14.4 & 1.86 & /\\
100\% & 2,970 & 230/230 (100.0\%) & 7.7 & 1.00 & /\\
\bottomrule
\end{tabular}

\end{table}

\subsection{External AS discrimination improves with local adaptation}\label{sec:2.6}

\noindent In a separate cross-cohort transfer analysis, AS classifiers developed in MIMIC-III were evaluated in EDU before any local adaptation. Zero-shot and adapted models were evaluated in the same 964 held-out patients, including 58 with high-grade AS, while 242 patients were reserved for local adaptation and validation. PulseAS--XGB achieved the highest zero-shot AUROC of 0.716, compared with 0.844 in the MIMIC-III source holdout. The end-to-end waveform model also retained discrimination, with a zero-shot AUROC of 0.705 (Table \ref{tab:4}).

Local adaptation strengthened this cross-cohort performance. Using approximately 20\% of eligible EDU patients for local adaptation increased the PulseAS--XGB AUROC to 0.813, with a paired gain of 0.097 (95\% CI, 0.058--0.137), while AP increased from 0.108 to 0.256. PulseAS--L2 improved from an AUROC of 0.646 to 0.805 and achieved the highest AP point estimate after adaptation (0.259). Gains were not confined to models incorporating clinical variables: PPG-only fusion improved from an AUROC of 0.663 to 0.771 and an AP of 0.097 to 0.224. All strategies undergoing parameter updates achieved higher AUROC and AP, whereas ExtraTrees, which underwent recalibration alone, retained its original ranking performance (Table \ref{tab:4}).

\begin{table}[!htbp]
\centering
\setcounter{table}{3}
\renewcommand{\theHtable}{main.\arabic{table}}
\caption{MIMIC-III-to-EDU AS validation and local adaptation}
\label{tab:4}
\fontsize{8}{10}\selectfont
\setlength{\tabcolsep}{3pt}
\renewcommand{\arraystretch}{1.0}
\noindent\begin{tabularx}{\textwidth}{@{}>{\raggedright\arraybackslash}X c c c c c c c@{}}
\toprule
\textbf{Model} & \multicolumn{3}{c}{\textbf{AUROC}} & \multicolumn{2}{c}{\textbf{AP}} & \multicolumn{2}{c}{\textbf{Sens@60\%Spec}} \\
\cmidrule(lr){2-4}\cmidrule(lr){5-6}\cmidrule(l){7-8}
& \makecell[c]{Zero-\\shot} & Adapted & $\Delta$ (95\% CI) & \makecell[c]{Zero-\\shot} & Adapted & \makecell[c]{Zero-\\shot} & Adapted \\
\midrule
Full pyPPG + ExtraTrees & 0.656 & 0.656 & \makecell[c]{0.000\\(0.000--0.000)} & 0.102 & 0.102 & 0.759 & 0.603 \\
Frozen AnyPPG + XGBoost & 0.631 & 0.745 & \makecell[c]{+0.114\\(0.073--0.155)} & 0.095 & 0.210 & 0.897 & 0.121 \\
PPG fusion + XGBoost & 0.663 & 0.771 & \makecell[c]{+0.108\\(0.064--0.152)} & 0.097 & 0.224 & 0.931 & 0.672 \\
PPG neural fusion & 0.657 & 0.764 & \makecell[c]{+0.108\\(0.030--0.187)} & 0.120 & 0.231 & 0.862 & 0.397 \\
\addlinespace[4pt]
Clinical + L2 logistic regression & 0.559 & 0.645 & \makecell[c]{+0.086\\(0.065--0.107)} & 0.064 & 0.095 & 0.793 & 0.414 \\
\addlinespace[4pt]
PulseAS--XGB & \textbf{0.716} & \textbf{0.813} & \makecell[c]{+0.097\\(0.058--0.137)} & 0.108 & 0.256 & 0.966 & 0.741 \\
PulseAS--MLP & 0.613 & 0.750 & \makecell[c]{+0.138\\(0.059--0.220)} & 0.088 & 0.214 & 0.966 & 0.534 \\
PulseAS--L2 & 0.646 & 0.805 & \makecell[c]{+0.159\\(0.088--0.233)} & 0.094 & 0.259 & 0.931 & 0.586 \\
\addlinespace[4pt]
End-to-end PPG & 0.705 & 0.783 & \makecell[c]{+0.078\\(0.031--0.124)} & 0.140 & 0.195 & 0.879 & 0.241 \\
\bottomrule
\end{tabularx}
\begin{tablenotes}[flushleft]
\fontsize{8}{10}\selectfont
\item[] Bold values indicate the highest AUROC point estimate in each evaluation setting.
\item[] $\Delta$AUROC = adapted minus zero-shot.
\end{tablenotes}
\end{table}

\section{Discussion}\label{sec:3}

\noindent PulseAS integrates deep PPG representations, waveform-derived biomarkers and routine clinical information to identify high-grade AS and characterise pulse changes around valve intervention. Disease-associated morphology, matched clinical comparisons and external validation provide complementary evidence for the information contained in short-duration PPG. Perioperative recordings extend this evidence to changes after treatment. Building on previous PPG-based valve-disease classification and representation learning\cite{PASS,PiLA,CardioPPG}, these findings link AS identification to an observable peripheral pulse phenotype and support accessible tools for pre-screening patients for confirmatory echocardiography and for repeated physiological assessment.

\noindent\needspace{3\baselineskip}
\noindent\textbf{Incremental value of PPG beyond routine clinical variables}\par

\noindent The matched clinical comparisons assessed the information contributed by PPG beyond routinely measured patient characteristics. Holding the classifier, patients and validation folds constant separated the addition of PPG from changes in modelling strategy. Higher AUROC point estimates after adding PPG were retained across alternative blood-pressure-based predictors, complete-case refitting and nonlinear clinical models, with more precise evidence for an increment in MIMIC-III. These findings support complementary predictive information beyond the measured demographic and blood-pressure variables. Importantly, these clinical parameters are routinely and easily obtained by paramedical staff during follow-up visits and can potentially be acquired by patients themselves using widely available devices. Unlike auscultation or ECG, which typically require dedicated trained personnel and specialised equipment, this combination of simple clinical variables and short-duration PPG offers a scalable approach that could facilitate risk stratification and earlier disease detection outside conventional cardiology settings. PPG could therefore augment routine clinical assessment with waveform information to support earlier AS detection and closer monitoring.

\noindent\needspace{3\baselineskip}
\noindent\textbf{AS-associated peripheral pulse morphology and perioperative changes}\par

\noindent Population waveforms and quantitative morphology features identified an AS-associated pattern involving pulse timing, systolic width and amplitude relationships. The later systolic peaking observed in both cohorts is consistent with the delayed arterial upstroke of AS\cite{ASWave2023}. Because PPG records the peripheral consequences of ventricular ejection and arterial transmission, this pattern reflects the interaction of cardiac and vascular physiology rather than a direct measurement of valve anatomy\cite{pyPPG,SVA2024}. The involvement of multiple waveform domains supports a multidimensional description of the peripheral expression of AS. Paired perioperative recordings extended this assessment to within-patient changes. Postoperative median pulses peaked earlier and showed narrower systolic contours, while longitudinal feature analysis revealed a higher diastolic-to-systolic width ratio. These changes were directionally consistent with the differences observed between comparison patients and those with high-grade AS in both cohorts. Although the incremental AUROC from adding PPG to routine clinical parameters was modest—likely reflecting partial overlap between waveform and clinical features—the clear, directionally consistent perioperative pulse changes provide strong physiological support for PPG-based AS identification and a basis for evaluating morphology during perioperative assessment and subsequent follow-up after TAVI.

\noindent\textbf{Potential for score-guided screening}\par

\noindent A PPG-assisted screening pathway could use PulseAS scores to prioritise patients for echocardiography. Supporting this approach, the highest-scoring tenth of each cohort contained more than two-fifths of its high-grade AS cases. In EDU, this stratum also captured more than half of the pre-TAVI cases. This concentration indicates that model scores relate both to disease severity and to recorded proximity to the need for intervention. Although the primary analysis included moderate AS and concomitant valve disease---placing the assessment in a clinically heterogeneous population---the intended screening target remains AS more severe than moderate. The present results therefore support evaluation of score-guided referral for confirmatory imaging, rather than using the score itself to determine an indication for valve intervention. Future prospective studies should establish the balance between case detection, missed disease, and the resulting echocardiographic workload.

\noindent\textbf{Cross-institutional transfer and adaptation}\par

\noindent Evaluation in Hong Kong and Boston assessed model transfer across institutions with different patient populations and clinical practices. The predominantly Asian EDU cohort and predominantly White MIMIC-III population span echocardiography attendance and critical care, providing geographically, demographically and clinically distinct settings. Source-trained PulseAS--XGB retained AS discrimination in EDU without local adaptation, and adaptation using 20\% of eligible target patients increased AUROC from 0.716 to 0.813 on the same held-out test set. This provides evidence of cross-institutional transfer across contrasting populations, while the further improvement after adaptation supports combining source-trained models with limited local data when extending PPG-based assessment to a new clinical setting.

\noindent\needspace{3\baselineskip}
\noindent\textbf{Limitations}\par

\noindent First, incomplete quantitative echocardiographic characterisation limits interpretation of the scores as measures of haemodynamic severity. Although the current evidence supports high-grade AS identification, a continuous relationship with valve obstruction and its cardiac consequences has not been established. Time-matched measurements of aortic valve area, peak velocity, mean gradient, stroke-volume index and ventricular structure and function, together with model-level attribution, would help link pulse features and learned representations to specific cardiac phenotypes. Previous work generating tissue Doppler waveforms from surface ECGs illustrates the value of paired physiological signals and imaging for such investigations\cite{ECGTDI}. The morphology findings provide physiological corroboration without resolving the mechanisms represented by the learned features.

Second, the longitudinal analysis was limited by the small follow-up sample and the absence of synchronised haemodynamic measurements. Changes in blood pressure, rhythm and perioperative treatment may accompany valve intervention, limiting attribution of the observed pulse changes to relief of obstruction alone. Continuous wearable PPG synchronised with invasive arterial pressure waveforms and procedural events could characterise the temporal correspondence between peripheral pulse changes and haemodynamic responses. Pilot work combining wrist PPG with invasive filling-pressure measurements supports the feasibility of this approach\cite{Scholte}. Repeated postoperative imaging and PPG, and longer follow-up in patients with established AS, would help distinguish immediate circulatory responses from longer-term adaptation and evaluate the value of individual pulse trajectories for prompting earlier reassessment.

Third, although the two-cohort design and external validation broaden the evidence base, both populations were hospital-based, and transfer was assessed in one source-to-target direction using retrospective partitions. Prospective multicentre studies should include more high-grade AS cases, expand demographic representation and evaluate both direct transfer and local adaptation in new patients. Differences in ethnic composition do not establish equivalent performance across all groups, and community screening requires evaluation in its intended population. Translation from fingertip recordings to ambulatory wearable monitoring will also require device- and site-specific assessment of waveform fidelity, signal quality and repeatability\cite{WristSQI} to establish the conditions and thresholds under which PPG can contribute to screening and follow-up.

\noindent\textbf{Conclusion}\par

\noindent Short-duration PPG combined with routine clinical data improved discrimination of high-grade AS and concentrated high-grade AS in higher AS-score strata. External discrimination and further gains with local adaptation support the potential of clinical–PPG models for AS-score stratification in new clinical populations, warranting prospective evaluation of their utility in prioritizing echocardiography.

\section{Methods}\label{sec:4}

\subsection{Study cohorts and reference standards}\label{sec:4.1}

\noindent Fingertip PPG and clinical data were obtained from Prince of Wales Hospital, Hong Kong (EDU), and MIMIC-III, linking its matched waveforms to ECHO-NOTE2NUM annotations\cite{MIMIC,MIMICWave,EchoNotes}. EDU recruitment included patients attending echocardiography and was not restricted to suspected AS. PPG and blood pressure were acquired during the same visit. Ethical approval was obtained from the Joint Chinese University of Hong Kong--New Territories East Cluster Clinical Research Ethics Committee (2026.578). Analyses used de-identified data.

AS grades were obtained from clinical echocardiography reports and classified according to the American Society of Echocardiography guidelines\cite{ASEAS2017}. High-grade AS comprised moderate-to-severe or severe disease in EDU. MIMIC-III used each patient’s highest valid historical AS grade, with severe disease classified as positive. Comparison patients had AS no greater than moderate, with concomitant valve disease retained. Patients with unknown AS grades were excluded. Unknown non-AS grades remained missing, and pulmonary valve grades were unavailable in MIMIC-III.

\subsection{PPG acquisition and representations}\label{sec:4.2}

\noindent EDU recordings were acquired at 60 Hz using a CONTEC CMS50E. Eligible 15-s segments required finite, nonconstant signals, stable amplitude and no contact-loss or abrupt-discontinuity artefacts, usable morphology features and a mean NeuroKit2 template-matching quality index \ensuremath{\geq}0.90\cite{NeuroKit}. MIMIC-III PPG recordings (PLETH channel) at 125 Hz were screened using MIMIC-III-Ext-PPG: all three 10-s sub-windows of each 30-s candidate had to pass\cite{MIMICWave,ExtPPG}. Accepted recordings were divided into 15-s segments, retaining at most 20 per patient by label-blind selection.

The 102 pyPPG variables were summarised within segments by mean, median, 25th and 75th percentiles, and interquartile range, yielding 510 morphology features\cite{pyPPG}. Adding standard deviation, skewness, kurtosis and median absolute deviation yielded 918 predictive candidates. Unreliable values were treated as missing. Inputs to the pretrained, frozen AnyPPG encoder were resampled to 125 Hz and standardised to zero mean and unit variance\cite{AnyPPG}. Training-fold selection retained ten features, at most one per morphology domain, ranked by \(r_j=\lvert\operatorname{AUROC}(f_j,y)-0.5\rvert\) for feature \(f_j\) and outcome \(y\).

Within PulseAS, frozen AnyPPG representations, selected morphology features and clinical variables were concatenated:

\begingroup
\setlength{\abovedisplayskip}{4pt plus 1pt minus 1pt}
\setlength{\belowdisplayskip}{4pt plus 1pt minus 1pt}
\setlength{\abovedisplayshortskip}{2pt plus 1pt minus 1pt}
\setlength{\belowdisplayshortskip}{2pt plus 1pt minus 1pt}
\postdisplaypenalty=10000
\begin{equation}\label{eq:fusion}
\mathbf z_{\mathrm{PPG}}=\left[f_{\theta_0}(x_{10});\,\mathbf m\right],
\qquad
\mathbf z_{\mathrm{clinical+PPG}}=\left[\mathbf z_{\mathrm{PPG}};\,\mathbf c\right].
\end{equation}
\endgroup

Here, \(f_{\theta_0}(x_{10})\in\mathbb R^{512}\) encodes the central 10-s window, \(\mathbf m\in\mathbb R^{10}\) contains training selected morphology features, and \(\mathbf c\) contains clinical variables and missingness indicators. Brackets denote concatenation.

\subsection{Model development and internal validation}\label{sec:4.3}

\noindent Model development and nested internal validation were conducted separately within EDU and MIMIC-III, with classifier fitting and data-dependent preprocessing confined to the training patients of the corresponding cohort. Eight strategies used morphology, AnyPPG representations, clinical inputs or their combinations, with ExtraTrees, XGBoost, L2-regularised logistic regression or a multilayer-perceptron head (Table~\ref{tab:2})\cite{ExtraTrees,XGBoost}. Clinical models used demographic and haemodynamic variables with missingness indicators. Patients with documented AF were retained in the analyses, but AF status and its missingness indicator were not used as predictors. MIMIC-III segments received equal total weight per patient. Segment scores were averaged within each patient before Platt calibration:

\begingroup
\setlength{\abovedisplayskip}{4pt plus 1pt minus 1pt}
\setlength{\belowdisplayskip}{4pt plus 1pt minus 1pt}
\setlength{\abovedisplayshortskip}{2pt plus 1pt minus 1pt}
\setlength{\belowdisplayshortskip}{2pt plus 1pt minus 1pt}
\postdisplaypenalty=10000
\begin{equation}\label{eq:patient-score}
\bar s_i=\frac{1}{m_i}\sum_{k=1}^{m_i}s_{ik},
\qquad
\hat p_i=\operatorname{logit}^{-1}\!\left(a+b\bar s_i\right).
\end{equation}
\endgroup

Here, \(m_i\) counts eligible segments and \(s_{ik}\) is segment \(k\)'s pre-calibration score. EDU scores came directly from patient-level features.

Nested validation used five patient-grouped outer folds and three inner folds shared across strategies, keeping all records from each patient together\cite{NestedCV}. Calibration parameters \(a\) and \(b\) were fitted using inner-training OOF patient scores\cite{Platt}. Feature selection, imputation, scaling and tuning used only training patients. Model selection prioritised AUROC, with AP resolving ties. Thresholds maximised sensitivity subject to training specificity \ensuremath{\geq}60\% and were applied unchanged to outer-test patients. Specificity in the test set was not fixed.

\subsection{Clinical comparisons and score stratification}\label{sec:4.4}

\noindent Matched L2 models compared five clinical baselines with and without PPG using identical patients and folds: DBP; pulse pressure (SBP\ensuremath{-}DBP); age plus either pressure measure; and age, sex, SBP and DBP. Sensitivity analyses refitted patients with observed age, sex, SBP and DBP and used restricted cubic splines with training-set knots at the 10th, 50th and 90th percentiles.

Selected-model OOF analyses stratified age, sex, observed SBP (\textless{}130/\ensuremath{\geq}130 mmHg), DBP (\textless{}80/\ensuremath{\geq}80 mmHg) and missing measurements. The SBP and DBP cutoffs correspond to the respective thresholds for hypertension in the 2017 American College of Cardiology/American Heart Association (ACC/AHA) guideline\cite{Whelton2018ACC2017}. Exploratory age bands were \textless{}50, 50--64, 65--74 and \ensuremath{\geq}75 years. For the restricted-control sensitivity analysis, comparison patients were required to have complete grading and no more than mild disease in all eight assessed valve lesions in EDU or all six in MIMIC-III. MIMIC-III used the highest valid historical grade for each lesion. All AS cases were retained, and performance was evaluated using the original OOF predictions and screening thresholds.

Patients were ranked retrospectively by PulseAS score, the calibrated OOF probability from each cohort’s highest-AUROC implementation (PulseAS--XGB in EDU and PulseAS--L2 in MIMIC-III), with higher scores indicating greater likelihood of high-grade AS. For \(N\) patients with \(E\) cases, the highest-scoring fraction \(f\) selected \(n_f\) patients and \(E_f\) cases:

\begingroup
\setlength{\abovedisplayskip}{4pt plus 1pt minus 1pt}
\setlength{\belowdisplayskip}{4pt plus 1pt minus 1pt}
\setlength{\abovedisplayshortskip}{2pt plus 1pt minus 1pt}
\setlength{\belowdisplayshortskip}{2pt plus 1pt minus 1pt}
\postdisplaypenalty=10000
\begin{equation}\label{eq:enrichment}
\operatorname{Capture}(f)=\frac{E_f}{E},
\qquad
\operatorname{Enrichment}(f)=\frac{E_f/n_f}{E/N}.
\end{equation}
\endgroup

Counts were rounded up with boundary ties included in \(n_f\). The case proportion in the selected group was \(E_f/n_f\). AS and pre-TAVI analyses used their respective case totals. EDU pre-TAVI cases required registry-confirmed procedures and pre-procedure index PPG. Pre-TAVI status was neither an input nor a prediction target. Capture for cases linked to later postoperative records used that subset’s denominator.

\subsection{Waveform and longitudinal analyses}\label{sec:4.5}

\noindent Complete pulses were onset-aligned, baseline-shifted and amplitude-normalised without time warping. Second derivatives used seconds-based units and the same amplitude scaling. Pointwise medians were aggregated across pulses, segments and patients on a 16.7-ms grid. Template points required \ensuremath{\geq}50\% pulse/segment support, and the common display interval required \ensuremath{\geq}90\% patient coverage in every group. Pointwise 95\% percentile intervals used 2,000 patient-bootstrap resamples, sharing patient draws across paired stages. Feature-level tests used individual observations.

For MIMIC-III, features were first median-aggregated within patients. Cross-sectional comparisons used asymptotic, continuity-corrected two-sided Mann--Whitney U tests and Cliff's delta, without imputation or outlier removal\cite{Cliff}.

\noindent\begin{minipage}[t]{\linewidth}
For AS feature \(y_{it}\) in patient \(i\) at stage \(t\), preTAVI--postTAVI differences used a patient random-intercept model:

\begingroup
\setlength{\abovedisplayskip}{4pt plus 1pt minus 1pt}
\setlength{\belowdisplayskip}{4pt plus 1pt minus 1pt}
\setlength{\abovedisplayshortskip}{2pt plus 1pt minus 1pt}
\setlength{\belowdisplayshortskip}{2pt plus 1pt minus 1pt}
\postdisplaypenalty=10000
\begin{equation}\label{eq:longitudinal}
y_{it}=\beta_0+\beta_{\mathrm{post}}\mathbf{1}(t=\mathrm{post})+u_i+\varepsilon_{it}.
\end{equation}
\endgroup

Here, \(u_i\) is a patient random intercept, \(\varepsilon_{it}\) is residual error, and \(\beta_{\mathrm{post}}\) estimates the postoperative--preoperative difference. Models used paired and unpaired observations and were fitted by restricted maximum likelihood, with two-sided Wald tests. Patient-cluster-robust models provided sensitivity analyses. Follow-up comparisons used paired Wilcoxon tests with automatic exact/asymptotic selection.
\end{minipage}\par

Benjamini--Hochberg correction covered all 510 features within each cohort--endpoint comparison or longitudinal contrast, retaining unestimable tests in the family. Statistical significance was defined as q\ensuremath{<}0.05\cite{BH}. Descriptive reference distributions comprised EDU patients with no more than mild disease in all eight assessed lesions.

\subsection{External validation and local adaptation}\label{sec:4.6}

\noindent To assess cross-cohort generalisability, AS classifiers developed and calibrated in MIMIC-III were evaluated in EDU, using consistent definitions of clinical variables and missingness indicators. Eligible EDU patients were stratified by AS status and divided into an adaptation and validation subset (20\%) and an independent test subset (80\%). Direct external validation (zero-shot evaluation) was performed without modifying the original models, preprocessing, calibration or decision thresholds.

We subsequently adapted the models using EDU data while retaining the features selected in MIMIC-III and the original preprocessing procedures where applicable. XGBoost models were updated by adding trees, with model selection based on EDU validation data, whereas logistic regression coefficients were updated with regularisation towards their original values. For neural fusion models, only the prediction heads were updated, with the AnyPPG encoder remaining frozen. The end-to-end waveform model underwent full-network fine-tuning, while ExtraTrees received recalibration alone. Calibration and threshold selection for the other adapted models were performed using EDU validation data under the original specificity criterion. The original and locally adapted models were evaluated on the same held-out EDU test set, which was excluded from all model development and adaptation procedures in this experiment. Pre-calibration discrimination was reported separately.

\subsection{Statistical analysis}\label{sec:4.7}

\noindent Patients were the independent analysis units. Clinical variables were summarised as medians [interquartile ranges] or available-case counts and percentages, with unadjusted two-sided Mann--Whitney U or Fisher exact tests. Performance measures were AUROC, AP, sensitivity, specificity, F1, balanced accuracy and Brier score. Reliability plots used ten approximately equal-sized groups.

Performance 95\% percentile CIs used 2,000 patient-bootstrap resamples of fixed predictions: ordinary resampling for internal EDU AS and outcome-stratified resampling for MIMIC-III AS. Blood-pressure subgroup CIs used outcome-stratified resampling in both cohorts. Paired comparisons shared patient draws, preserving outcome, age, control-eligibility or case-membership strata as applicable. Four-band age-group CIs used 20,000 age-by-AS-stratified resamples. Model-difference intervals were unadjusted. Models were not refitted, so the intervals excluded model-selection uncertainty. Analyses used Python, SciPy, statsmodels, scikit-learn, XGBoost, PyTorch, NeuroKit2 and pyPPG.

\begingroup
\section*{Data availability}

\noindent The individual-level data from the EDU cohort are not publicly available owing to institutional ethics restrictions and hospital data governance policies. Access to de-identified EDU data may be considered upon reasonable request to the corresponding author, subject to approval by the relevant ethics committee and institution.

The MIMIC-III Clinical Database, MIMIC-III Waveform Database Matched Subset and ECHO-NOTE2NUM used in this study are available through PhysioNet under their respective access policies\cite{MIMIC,MIMICWave,EchoNotes}. Access to the MIMIC-III clinical data and ECHO-NOTE2NUM requires PhysioNet credentialing, completion of the required training and acceptance of the relevant data use agreements. The matched waveform subset is openly available under the Open Data Commons Open Database License v1.0.

\section*{Code availability}

\noindent The core analysis code for PulseAS is publicly available at \url{https://github.com/Joseph-wang-sus/PulseAS}. Net1D used the \href{https://github.com/hsd1503/resnet1d}{resnet1d software implementation by Shenda Hong}. AnyPPG weights are obtained under the upstream license. Model-access and representation-generation instructions are provided in the repository.
\endgroup

\clearpage
\begingroup
\setlength{\bibsep}{0.85em}
\interlinepenalty=10000

\endgroup
\section*{Acknowledgements}

\noindent We acknowledge the staff at EDU for assisting the PPG collection.

\section*{Author contributions}

\noindent Haowan Wang conceived and designed the study, implemented the algorithms, analyzed the results, and drafted the manuscript. Cheng Ding contributed to methodological discussions and manuscript revision. Donglin Xie contributed to data curation, preprocessing, supporting analyses, and manuscript revision. Huizi Yu and Guangkun Nie contributed to software development, validation, and formal analysis. Bryan P Yan, Shenda Hong, and Kent Chak-yu So supervised the project, provided overall technical guidance, and critically reviewed and revised the manuscript. All authors reviewed and approved the final version of the manuscript.

\section*{Funding}

\noindent The project is funded by the Chinese University of Hong Kong Start-up Fund for Prof So Chak Yu (4937241).

\section*{Competing interests}

\noindent The authors declare no competing interests.

\section*{Consent to Participate declaration}

\noindent Not applicable.
\clearpage

\setcounter{figure}{0}
\setcounter{table}{0}
\renewcommand{\thefigure}{S\arabic{figure}}
\renewcommand{\thetable}{S\arabic{table}}
\renewcommand{\theHfigure}{supp.\arabic{figure}}
\renewcommand{\theHtable}{supp.\arabic{table}}
\renewcommand{\figurename}{Supplementary Fig.}
\renewcommand{\tablename}{Supplementary Table}

\section*{Supplementary Information}
\setlength{\parindent}{0pt}

\Needspace{18\baselineskip}
\subsection*{AS Supplementary Evidence}

\begin{figure}[!htbp]
\centering

\setbox0=\vbox{\hsize=\linewidth\small\noindent\textbf{\figurename~\thefigure.} Waveform origins of representative AS morphology features. \textbf{a,} An illustrative PPG pulse and its independently rescaled first and second derivatives, with selected timing, width and amplitude landmarks. \textbf{b,} Waveform interpretations of the ten representative features. AI denotes augmentation index, distinct from the aging index.\par}
\includegraphics[width=\linewidth,height=\dimexpr\textheight-\ht0-\dp0-42pt\relax,keepaspectratio]{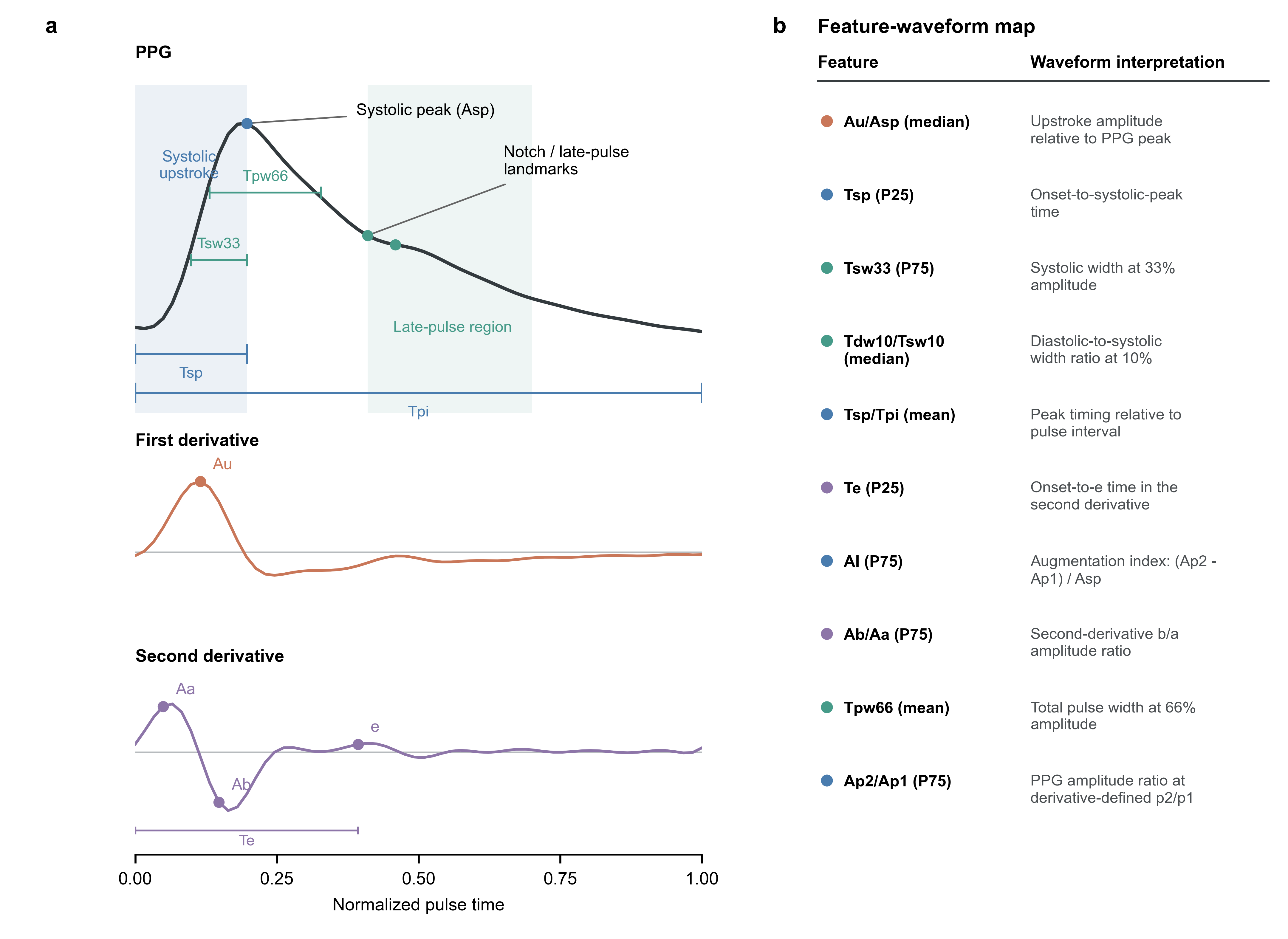}
\caption{Waveform origins of representative AS morphology features. \textbf{a,} An illustrative PPG pulse and its independently rescaled first and second derivatives, with selected timing, width and amplitude landmarks. \textbf{b,} Waveform interpretations of the ten representative features. AI denotes augmentation index, distinct from the aging index.}\label{fig:S1}
\end{figure}
\FloatBarrier

\begin{figure}[!htbp]
\centering

\setbox0=\vbox{\hsize=\linewidth\small\noindent\textbf{\figurename~\thefigure.} AS probability calibration. \textbf{a,b,} EDU and MIMIC-III. Points show observed case proportions against mean predicted probability in ten approximately equal-sized groups. Clinical-only and PPG models use L2 logistic regression and fusion XGBoost, respectively; PulseAS--XGB and PulseAS--L2 are shown for EDU and MIMIC-III, respectively. Dashed diagonals indicate ideal calibration.\par}
\includegraphics[width=\linewidth,height=\dimexpr\textheight-\ht0-\dp0-42pt\relax,keepaspectratio]{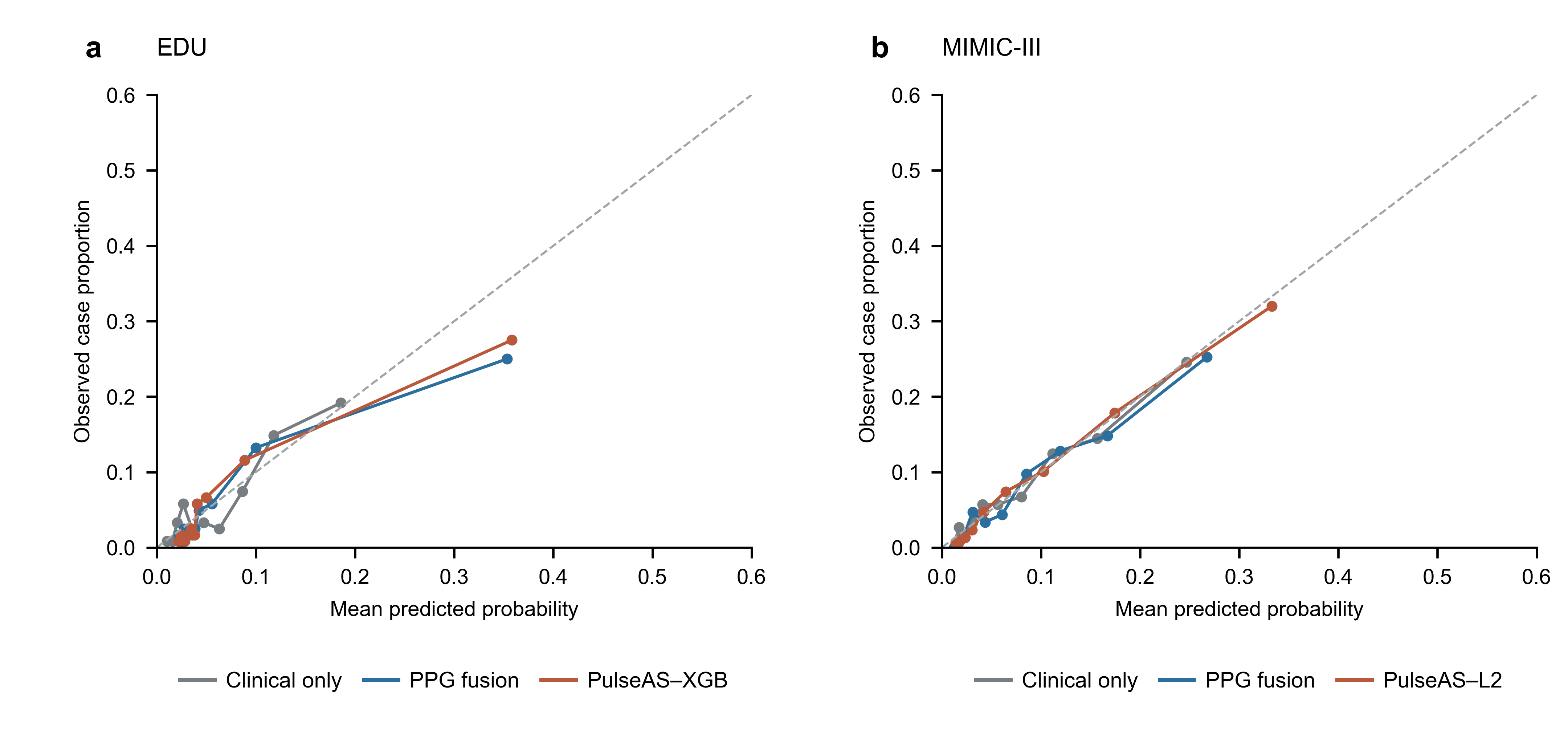}
\caption{AS probability calibration. \textbf{a,b,} EDU and MIMIC-III. Points show observed case proportions against mean predicted probability in ten approximately equal-sized groups. Clinical-only and PPG models use L2 logistic regression and fusion XGBoost, respectively; PulseAS--XGB and PulseAS--L2 are shown for EDU and MIMIC-III, respectively. Dashed diagonals indicate ideal calibration.}\label{fig:S2}
\end{figure}
\FloatBarrier

\begingroup
\Needspace{8\baselineskip}

\setcounter{table}{0}
\refstepcounter{table}\label{tab:S1}

\par\noindent{\normalfont\normalsize\bfseries Supplementary Table S1 \textbar{} AS classification with low-valvular-disease burden controls\par}\nobreak\smallskip

\begingroup
\fontsize{8}{10.5}\selectfont
\setlength{\tabcolsep}{3pt}
\renewcommand{\arraystretch}{1.2}
\stepcounter{longtableblock}
\renewcommand{\theHtable}{block.\arabic{longtableblock}}
\addtocounter{table}{-1}
\begin{longtable}{>{\raggedright\arraybackslash}p{164.78767pt}rrrrr}
\toprule
Model & \shortstack[r]{AUROC (95\% CI)} & \shortstack[r]{AP (95\% CI)} & \shortstack[r]{Sensitivity} & \shortstack[r]{Specificity} & \shortstack[r]{Balanced\\accuracy}\\
\midrule
\endfirsthead
\multicolumn{6}{l}{Supplementary Table \ref{tab:S1} (continued)}\\
\toprule
Model & \shortstack[r]{AUROC (95\% CI)} & \shortstack[r]{AP (95\% CI)} & \shortstack[r]{Sensitivity} & \shortstack[r]{Specificity} & \shortstack[r]{Balanced\\accuracy}\\
\midrule
\endhead
\bottomrule
\endfoot
\bottomrule
\endlastfoot
\multicolumn{6}{l}{\textbf{EDU}}\\*
Full pyPPG + ExtraTrees & 0.822 (0.761--0.879) & 0.433 (0.337--0.553) & 0.822 & 0.678 & 0.750\\
Clinical + Logistic L2 & 0.760 (0.704--0.818) & 0.218 (0.173--0.303) & 0.712 & 0.678 & 0.695\\
Frozen AnyPPG + XGBoost & 0.822 (0.767--0.876) & 0.409 (0.318--0.523) & 0.849 & 0.574 & 0.712\\
AnyPPG--Top-10 pyPPG fusion + XGBoost & 0.811 (0.755--0.864) & 0.377 (0.287--0.502) & 0.863 & 0.541 & 0.702\\
AnyPPG--Top-10 neural fusion & 0.751 (0.685--0.817) & 0.349 (0.255--0.460) & 0.822 & 0.600 & 0.711\\
PulseAS--XGB & 0.834 (0.784--0.885) & 0.448 (0.349--0.562) & 0.863 & 0.575 & 0.719\\
PulseAS--MLP & 0.768 (0.701--0.835) & 0.371 (0.277--0.489) & 0.781 & 0.640 & 0.710\\
PulseAS--L2 & 0.812 (0.756--0.866) & 0.407 (0.313--0.522) & 0.822 & 0.707 & 0.764\\
\midrule
\multicolumn{6}{l}{\textbf{MIMIC-III}}\\*
Full pyPPG + ExtraTrees & 0.828 (0.799--0.855) & 0.499 (0.444--0.560) & 0.826 & 0.719 & 0.772\\
Clinical + Logistic L2 & 0.851 (0.824--0.876) & 0.570 (0.512--0.631) & 0.774 & 0.785 & 0.780\\
Frozen AnyPPG + XGBoost & 0.835 (0.806--0.862) & 0.502 (0.448--0.568) & 0.857 & 0.669 & 0.763\\
AnyPPG--Top-10 pyPPG fusion + XGBoost & 0.838 (0.809--0.863) & 0.506 (0.450--0.570) & 0.839 & 0.684 & 0.762\\
AnyPPG--Top-10 neural fusion & 0.836 (0.807--0.864) & 0.512 (0.460--0.576) & 0.861 & 0.665 & 0.763\\
PulseAS--XGB & 0.899 (0.878--0.918) & 0.664 (0.607--0.724) & 0.913 & 0.728 & 0.821\\
PulseAS--MLP & 0.876 (0.851--0.899) & 0.625 (0.566--0.686) & 0.883 & 0.693 & 0.788\\
PulseAS--L2 & 0.897 (0.875--0.916) & 0.665 (0.607--0.723) & 0.896 & 0.747 & 0.821\\
\end{longtable}
\endgroup

\endgroup

\begingroup
\Needspace{8\baselineskip}

\setcounter{table}{1}
\refstepcounter{table}\label{tab:S2}

\par\noindent{\normalfont\normalsize\bfseries Supplementary Table S2 \textbar{} Clinical baselines and incremental AS information from PPG\par}\nobreak\smallskip

\begingroup
\fontsize{8}{10.5}\selectfont
\setlength{\tabcolsep}{3pt}
\renewcommand{\arraystretch}{1.2}
\stepcounter{longtableblock}
\renewcommand{\theHtable}{block.\arabic{longtableblock}}
\addtocounter{table}{-1}
\begin{longtable}{>{\raggedright\arraybackslash}p{241.83166pt}rrr}
\toprule
Clinical baseline / analysis & \shortstack[r]{Clinical-only\\AUROC} & \shortstack[r]{Clinical + PPG\\AUROC} & \shortstack[r]{\ensuremath{\Delta}AUROC (95\% CI)}\\
\midrule
\endfirsthead
\multicolumn{4}{l}{Supplementary Table \ref{tab:S2} (continued)}\\
\toprule
Clinical baseline / analysis & \shortstack[r]{Clinical-only\\AUROC} & \shortstack[r]{Clinical + PPG\\AUROC} & \shortstack[r]{\ensuremath{\Delta}AUROC (95\% CI)}\\
\midrule
\endhead
\bottomrule
\endfoot
\bottomrule
\endlastfoot
\multicolumn{4}{l}{\textbf{EDU}}\\*
DBP & 0.711 & 0.781 & 0.070 (-0.003 to 0.141)\\
Pulse pressure & 0.578 & 0.767 & 0.189 (0.108 to 0.264)\\
Age + DBP & 0.747 & 0.795 & 0.048 (-0.019 to 0.108)\\
Age + pulse pressure & 0.715 & 0.783 & 0.068 (-0.007 to 0.140)\\
Age + sex + SBP + DBP & 0.744 & 0.801 & 0.057 (-0.015 to 0.123)\\
Clinical-4, complete cases (linear) & 0.714 & 0.782 & 0.068 (-0.012 to 0.145)\\
Clinical-4, complete cases (nonlinear) & 0.712 & 0.788 & 0.076 (-0.005 to 0.153)\\
\midrule
\multicolumn{4}{l}{\textbf{MIMIC-III}}\\*
DBP & 0.647 & 0.813 & 0.165 (0.128 to 0.202)\\
Pulse pressure & 0.640 & 0.811 & 0.172 (0.133 to 0.212)\\
Age + DBP & 0.763 & 0.838 & 0.075 (0.050 to 0.101)\\
Age + pulse pressure & 0.760 & 0.837 & 0.077 (0.052 to 0.103)\\
Age + sex + SBP + DBP & 0.760 & 0.837 & 0.077 (0.052 to 0.104)\\
Clinical-4, complete cases (linear) & 0.732 & 0.824 & 0.092 (0.060 to 0.127)\\
Clinical-4, complete cases (nonlinear) & 0.731 & 0.821 & 0.089 (0.058 to 0.124)\\
\end{longtable}
\endgroup

\endgroup

\begin{figure}[!htbp]
\centering

\setbox0=\vbox{\hsize=\linewidth\small\noindent\textbf{\figurename~\thefigure.} AS-associated effects across waveform domains. \textbf{a,b,} EDU and MIMIC-III, respectively. Each point represents one of 510 morphology features. Cliff's delta compares AS-positive with comparison patients; colors indicate Benjamini--Hochberg q values corrected within each cohort.\par}
\includegraphics[width=\linewidth,height=\dimexpr\textheight-\ht0-\dp0-42pt\relax,keepaspectratio]{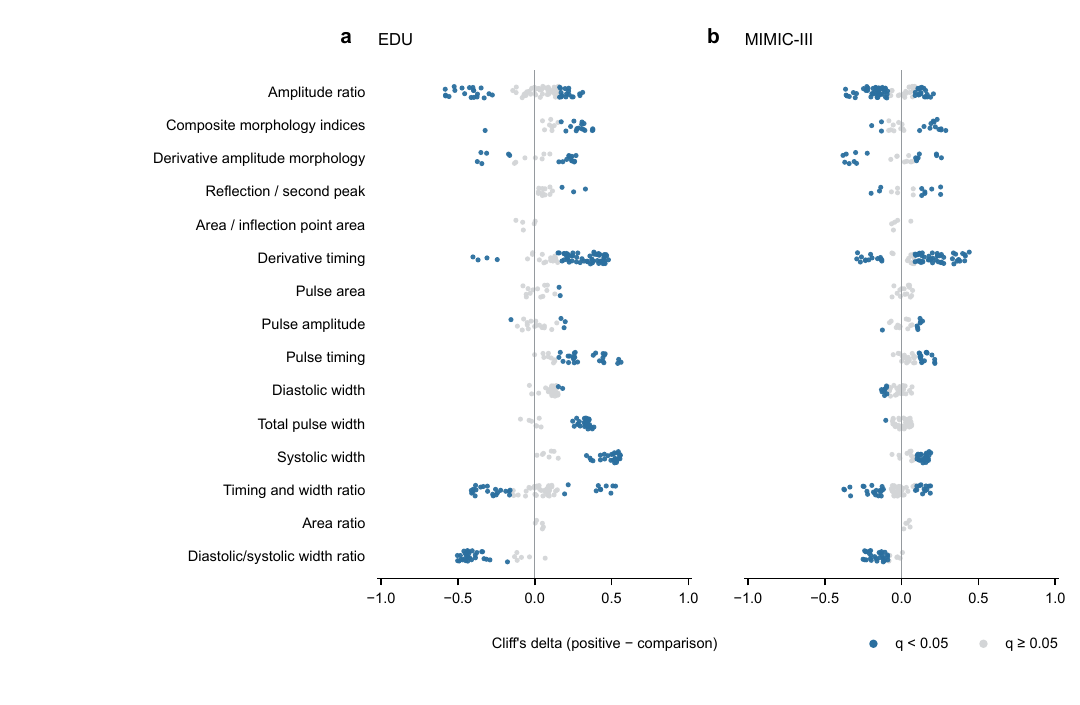}
\caption{AS-associated effects across waveform domains. \textbf{a,b,} EDU and MIMIC-III, respectively. Each point represents one of 510 morphology features. Cliff's delta compares AS-positive with comparison patients; colors indicate Benjamini--Hochberg q values corrected within each cohort.}
\label{fig:S3}
\end{figure}
\FloatBarrier

\begin{figure}[!htbp]
\centering

\setbox0=\vbox{\hsize=\linewidth\small\noindent\textbf{\figurename~\thefigure.} AS discrimination across four age groups. \textbf{a,b,} PulseAS--XGB and PulseAS--L2 in EDU and MIMIC-III, respectively. Points and bars indicate AUROC and 95\% patient-bootstrap CIs from fixed OOF predictions; o denotes groups with fewer than ten AS-positive patients. Dashed lines indicate chance discrimination.\par}
\includegraphics[width=\linewidth,height=\dimexpr\textheight-\ht0-\dp0-42pt\relax,keepaspectratio]{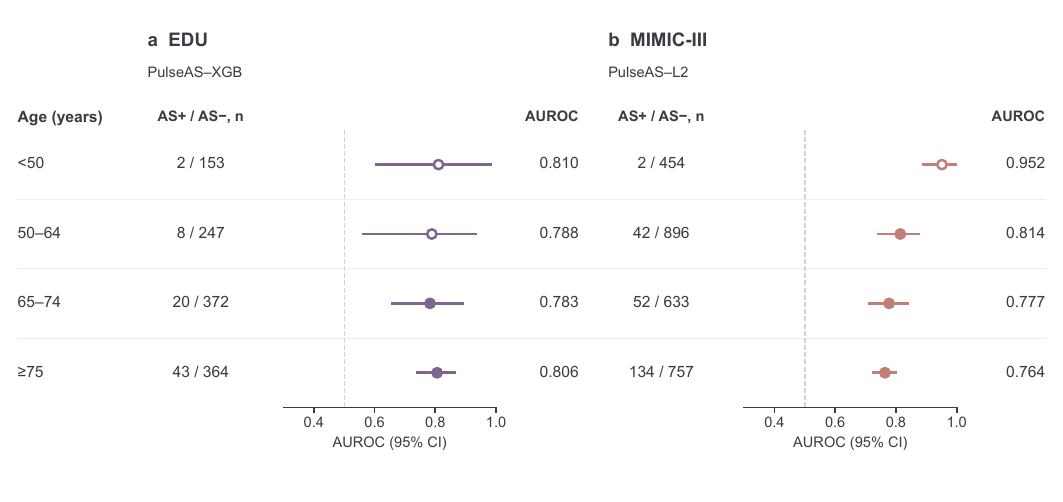}
\caption{AS discrimination across four age groups. \textbf{a,b,} PulseAS--XGB and PulseAS--L2 in EDU and MIMIC-III, respectively. Points and bars indicate AUROC and 95\% patient-bootstrap CIs from fixed OOF predictions; o denotes groups with fewer than ten AS-positive patients. Dashed lines indicate chance discrimination.}
\label{fig:S4}
\end{figure}
\FloatBarrier
\end{document}